\documentclass[twocolumn]{openjournal}
\usepackage[colorlinks,linkcolor=red,citecolor=blue,urlcolor=blue ]{hyperref}
\usepackage{amsmath,amssymb}
\usepackage{graphicx}
\usepackage{dcolumn}
\usepackage{bm}
\usepackage[dvipsnames]{xcolor}
\usepackage{hyperref}
\usepackage{orcidlink}
\usepackage{aas_macros}
\usepackage{enumitem}
\usepackage{comment}
\usepackage{booktabs,array}
\usepackage{multirow}
\usepackage{tabularx}

\begin{document}

\title{Accurate modeling for 3$\times$2pt analyses in Roman and Rubin: a study of model approximations\vspace{-1cm}}
\author{Junzhou Zhang$^{* \ 1}$\orcidlink{0009-0000-3461-4041}}
\author{Chihway Chang$^{2,3,4}$\orcidlink{0000-0002-7887-0896}}
\author{Jiachuan Xu$^{5}$\orcidlink{0000-0003-0871-8941}}
\author{Vivian Miranda$^{6}$\orcidlink{0000-0003-4776-0333}}
\author{Chun-Hao To$^{2}$\orcidlink{0000-0001-7836-2261}}
\author{Haley Bowden$^{7}$\orcidlink{0009-0006-1916-8550}}
\author{Kaili Cao$^{8,9}$\orcidlink{0000-0002-1699-6944}}
\author{Tim Eifler$^{7}$\orcidlink{0000-0002-1894-3301}}
\author{Roman HLIS Cosmology PIT}
\email{$^*$email: junzhou@uchicago.edu}

\affiliation{\\ \ \ \ \ $^{1}$ Department of Physics, University of Chicago, Chicago, IL 60637, USA}
\affiliation{$^{2}$ Department of Astronomy and Astrophysics, University of Chicago, Chicago, IL 60637, USA}
\affiliation{$^{3}$ Kavli Institute for Cosmological Physics, University of Chicago, Chicago, IL 60637, USA}
\affiliation{$^{4}$ NSF-Simons AI Institute for the Sky (SkAI), 172 E. Chestnut St., Chicago, IL 60611, USA}
\affiliation{$^{5}$ Northeastern University, Boston, MA, 02115, USA}
\affiliation{$^{6}$ C. N. Yang Institute for Theoretical Physics, Stony Brook University, Stony Brook, NY 11794, USA}
\affiliation{$^{7}$ Department of Astronomy/Steward Observatory, University of Arizona, 933 North Cherry Avenue, Tucson, AZ 85721-0065, USA}
\affiliation{$^{8}$ Center for Cosmology and AstroParticle Physics (CCAPP), The Ohio State University, 191 West Woodruff Ave, Columbus, OH 43210, USA}
\affiliation{$^{9}$ Department of Physics, The Ohio State University, 191 West Woodruff Ave, Columbus, OH 43210, USA}

\begin{abstract}
One of the pillars of modern cosmology is the use of galaxy imaging surveys to extract information from the large-scale structure. In recent surveys, this measurement is typically performed through a 3$\times$2pt analysis, which combines auto- and cross-correlations between galaxy density and galaxy weak lensing. In this paper, we carry out a systematic study of three modeling approximations commonly used in such analyses: 1) applying the Limber approximation, 2) neglecting redshift-space distortions, and 3) using less accurate models for the nonlinear matter power spectrum. We carry out the study in the context of the final data from two major Stage-IV galaxy imaging surveys: the Nancy Grace Roman Space Telescope's High Latitude Imaging Survey and the Vera C. Rubin Observatory's Legacy Survey of Space and Time. To do this, we first validate our modeling pipeline, implemented in the software package \textsc{CoCoA}, against an established code base, \textsc{CCL}. Next, we perform a simulated likelihood analysis to assess the impact of these approximations on the cosmological constraints. We find all three effects to be important; neglecting any of them can induce biases in cosmological constraints approaching or exceeding $1\sigma$, and exceeding $2\sigma$ for Rubin in several cases.
Moreover, we explore how the lens-galaxy sample configuration and scale-cut choice can influence the constraints.
\end{abstract}
\maketitle

\section{Introduction}
\label{sec:intro}
\vspace{0.2cm}
\makeatletter
\AtBeginDocument{
  \typeout{columnwidth=\the\columnwidth}
  \typeout{textwidth=\the\textwidth}
}
\makeatother

Galaxy imaging surveys provide one of the primary avenues for testing the standard $\Lambda$CDM cosmological paradigm through measurements of large-scale structure. Over the past two decades, the community has developed a mature framework for extracting this information from galaxy imaging data by combining three two-point correlation functions \citep{2004PhRvD..70d3009H, Mandelbaum_2018, Weinberg_2013}: the autocorrelation of weak-lensing shear (cosmic shear), the autocorrelation of galaxy density (galaxy clustering), and the cross-correlation between galaxy density and weak-lensing shear (galaxy--galaxy lensing). This joint analysis, commonly referred to as ``$3\times2$pt,'' has been performed by all major Stage-III galaxy imaging surveys \citep{Heymans2021_KiDS1000_3x2pt, Zhang2025_HSC3x2pt, DESY63x2pt2026}.

Combining these three complementary probes has several advantages. 
First, these probes depend on different combinations and powers of cosmological and nuisance parameters. For example, in a simplified linear-bias picture, the amplitudes of cosmic shear, galaxy--galaxy lensing, and galaxy clustering scale approximately as $\sigma_8^2$, $b\sigma_8^2$, and $b^2\sigma_8^2$, respectively, where $b$ is the linear galaxy bias. Their joint analysis can therefore break degeneracies between the amplitude of matter fluctuations ($\sigma_8$) and galaxy bias ($b$), providing substantially greater constraining power than any individual probe alone. Second, the three probes respond differently to astrophysical and observational systematics. Jointly analyzing them can therefore self-calibrate some of the associated nuisance parameters. In addition, requiring all three probes to be described consistently by a common cosmological and systematic model provides a powerful internal consistency test: residual systematic effects that primarily affect one probe may produce tension with the others and motivate further investigation. The recent Dark Energy Survey Year 6 (DES-Y6) $3\times2$pt analysis \citep{DESY63x2pt2026} illustrates these advantages, yielding tighter constraints than the individual probes while remaining stable across a broad range of modeling and analysis choices.

As Stage-III surveys are concluding, we are now facing new challenges from next-generation galaxy imaging surveys such as the Nancy Grace Roman Space Telescope's High Latitude Imaging Survey\footnote{\url{https://roman-hlis-cosmology.caltech.edu/}} (HLIS), the Rubin Observatory's Legacy Survey of Space and Time\footnote{\url{https://rubinobservatory.org/}} (LSST), and the Euclid mission\footnote{\url{https://www.esa.int/Science_Exploration/Space_Science/Euclid}}. These surveys are expected to provide unprecedented statistical power, which places stringent requirements on the control of systematic effects \citep{Eifler_2021, thelsstdarkenergysciencecollaboration2021lsstdarkenergyscience, 2020euclidforecast}. Therefore, it is important to re-examine all aspects of the 3$\times$2pt analysis and ensure that the pipelines used for Stage-III surveys remain valid. 

In this work, we focus on three specific modeling components in the 3$\times$2pt analysis. First, we investigate the impact of the Limber approximation in the modeling of observables. Second, we examine the impact of neglecting the redshift-space distortions (RSD) modeling. Finally, we look at how the 3$\times$2pt results are sensitive to the choice of the nonlinear matter power spectrum model. Several studies have investigated these modeling components. For the Limber approximation, \citet{Leonard_2023} showed that the usual first-order Limber approximation is insufficiently accurate for the full 10-year (Y10) LSST dataset on $\ell=200-1000$. \citet{Fang_2020} showed that using the non-Limber model can effectively correct the biases on constraints induced by the Limber approximation. They also showed that a full non-Limber model of galaxy clustering is necessary for the first year (Y1) of LSST 3$\times$2pt analysis to avoid significant bias. For the nonlinear power spectrum, \citet{sanchezcid2026darkenergysurveyyear} showed that using \textsc{HMCode} instead of \textsc{EuclidEmulator2} results in a negligible shift in cosmology when applied to the DES-Y6 3$\times$2pt analysis. In \citet{Tan_2023}, the authors suggest that the same conclusion holds for cosmic-shear-only analyses in stage-III surveys, whereas it becomes important for the Stage-IV surveys. The latter conclusion is also supported by \citet{Martinelli_2021}. For RSD, \citet{euclid_prep_XXXIV2023} showed that when the linear RSD is neglected, significant biases occur in $\Lambda$CDM with the final 6-year Euclid dataset. 

The main goal of this paper is to provide a systematic study of these three effects in the context of a 3$\times$2pt analysis for the final 5-year (Y5) Roman and 10-year (Y10) Rubin datasets. In addition, we also achieve two secondary goals. First, we update some of the analysis choices that incorporate our knowledge from Stage-III surveys. Second, we conduct a code comparison between the official modeling packages used by the Roman HLIS Cosmology PIT and the Rubin LSST Dark Energy Science Collaboration (DESC), namely \textsc{CoCoA} and \textsc{CCL}, respectively, to ensure the robustness of modeling frameworks across the Stage-IV surveys. 

The paper is structured as follows. In Section~\ref{sec:model-and-approx}, we briefly describe our modeling framework for the 3$\times$2pt data vector and highlight the three approximations we focus on. In Section~\ref{sec:surveys}, we describe the data configurations used for both Roman and Rubin 3$\times$2pt analyses. In Section~\ref{sec:likelihood}, we introduce the details of our simulated likelihood analysis. In Section~\ref{sec:results} we present our results and recommendations for the two surveys. We finally conclude in Section~\ref{sec:conclusion}. 

\vspace{0.3cm}
\section{Modeling and Approximations}
\label{sec:model-and-approx}
\vspace{0.2cm}
We first describe the overall framework for modeling the 3$\times$2pt data vector in Section~\ref{sec:model} and then focus on describing the three approximations we study in this work in Section~\ref{sec:approx}. 

In all our tests, we consider a $\Lambda$CDM model, which assumes dark energy is a cosmological constant and has five free parameters ($A_s$, $n_s$, $H_0$, $\Omega_\mathrm{b}$, $\Omega_\mathrm{m}$). Here, $A_s$ and $n_s$ are the amplitude and scalar spectral index of the primordial density fluctuation, $H_0$ is the Hubble constant, and $\Omega_\mathrm{m}$ and $\Omega_\mathrm{b}$ are the matter and baryon density parameters.

\subsection{Modeling 3$\times$2pt data vector}
\label{sec:model}
\vspace{0.2cm}
\subsubsection{Power spectrum}
\vspace{0.2cm}
We follow the formalism used in the DES-Y6 analysis \citep{sanchezcid2026darkenergysurveyyear} for modeling the 3$\times$2pt data vector. We focus on two projected fields: the projected galaxy overdensity field $(\delta_g)$ and the weak lensing shear field $(\gamma = \gamma_1+i\gamma_2)$, probed by the lens-galaxy and source-galaxy samples, respectively. For a given observation direction $\boldsymbol{\hat{n}}$, they are described below without correction terms for observational or astrophysical effects. For example, the contribution from intrinsic alignment to the observed weak lensing shear field is not included in the expression. 
\begin{align}
    \delta_{g,\mathrm{obs}}^i &= \delta_{g,D}^i(\boldsymbol{\hat{n}})= \int d\chi W_{\delta,g}^i \delta_g^{(3D)}(\boldsymbol{\hat{n}}\chi,\chi),\label{galaxy-overdensity-field}\\
    \gamma_1^i &= \gamma_{1,G}^i(\boldsymbol{\hat{n}}) = \int d\chi W_{\kappa,g}^i \chi (\Phi_{,11}-\Phi_{,22})(\boldsymbol{\hat{n}}\chi,\chi),\label{gravitational-shear-field-1}\\
    \gamma_2^i &= \gamma_{2,G}^i(\boldsymbol{\hat{n}}) = \int d\chi W_{\kappa,g}^i \chi (\Phi_{,12}+\Phi_{,21})(\boldsymbol{\hat{n}}\chi,\chi).\label{gravitational-shear-field-2}
\end{align}

The subscript $(g)$ represents the galaxy samples, which can denote either lens samples $(l)$ or source samples $(s)$. They are divided into different tomographic bins, and the superscript $(i)$ denotes the bin indices. $\chi$ is the comoving distance. $\delta_{g,D}$ denotes the line-of-sight projection component of $\delta_g^{(3D)}$, the three-dimensional galaxy density contrast field. $\gamma_{\alpha,G}$ ($\alpha=1,2$) denotes the gravitational shear components, derived from $\Phi_{,\alpha}$, the spatial transverse derivative of the $3D$ potential $\Phi$.
$W_{\delta,g}^i$ and $W_{\kappa, g}^i$ are the normalized selection function and the lensing efficiency, respectively. They are defined as,
\begin{align}
    W_{\delta,g}^i &= n_g^i(z(\chi))\frac{dz}{d\chi}, \label{selection function}\\
    W_{\kappa,g}^i &= \frac{3\Omega_\mathrm{m} H_0^2}{2c^2}\int_\chi^\infty d\chi' n_g^i(\chi')\frac{\chi}{a(\chi)}\frac{\chi'-\chi}{\chi'},\label{lensing efficiency}
\end{align}
where $z$ is the redshift, $a$ is the scale factor, and $n_g^i(z)$ is the normalized redshift distribution of the galaxy sample in tomographic bin $i$. 

We can then express the angular power spectrum of the two projected fields by computing the second moment of their spherical harmonic coefficients. For example, the angular power spectrum of two projected galaxy overdensity fields is,
\begin{align}
    \label{angular-power-spectrum-gg}
    C_{\ell, \delta_{g,D}\delta_{g,D}}^{ij} 
    &= \langle a_{\ell m}^{i} a_{\ell m}^{j*} \rangle\nonumber\\
    &= (4\pi)^2 \int d\chi_1 W_{\delta,g}^i(\chi_1)\int d\chi_2 W_{\delta,g}^j(\chi_2) \nonumber\\
    &\cdot\int \frac{d^3k_1}{(2\pi)^3}j_\ell(k_1\chi_1)Y_{\ell m }(\boldsymbol{\hat{k}_1})\nonumber\\
    &\cdot\int\frac{d^3k_2}{(2\pi)^3}j_\ell(k_2\chi_2)Y_{\ell m }^*(\boldsymbol{\hat{k}_2})\nonumber\\
    &\cdot\langle \tilde\delta_g^{(3D)}(\boldsymbol{k_1},\chi_1)\tilde\delta_g^{*(3D)}(\boldsymbol{k_2}, \chi_2) \rangle \nonumber\\
    &= \frac{2}{\pi} \int d\chi_1 W_{\delta,g}^i(\chi_1)\int d\chi_2 W_{\delta,g}^j(\chi_2) \nonumber\\
    &\cdot \int k^2 dk j_\ell(k\chi_1)j_\ell(k\chi_2)P_{gg}(\boldsymbol{k}, \chi_1,\chi_2).
\end{align}
where $a_{\ell m}$ is the coefficient of the spherical harmonic expansion and $\tilde\delta_g^{(3D)}$ is the galaxy overdensity field in Fourier space. In deriving the last line, we use the relation,
\begin{align}
    \langle \tilde\delta_g^{(3D)}(\boldsymbol{k_1},\chi_1), \tilde\delta_g^{(3D)}(\boldsymbol{k_2}, \chi_2) \rangle &= (2\pi)^3\cdot\delta_D(\boldsymbol{k_1} - \boldsymbol{k_2})\nonumber\\&\cdot P_{gg}(\boldsymbol{k_1},\chi_1,\chi_2).
\end{align}
Here $\delta_D$ is the Dirac delta function, $j_\ell$ is the spherical Bessel function, and $P_{gg}$ is the unequal time galaxy power spectrum. The superscripts $i$ and $j$ denote the tomographic-bin indices, while $\chi_1$, $\chi_2$, $\boldsymbol{k}_1$, and $\boldsymbol{k}_2$ are variables to distinguish the two line-of-sight integrals. The angular power spectra of cosmic shear and galaxy-galaxy lensing can be derived using the same procedure.

\subsubsection{Correlation Functions}
\label{sec:model-correlation function}
\vspace{0.2cm}
Correlation functions in real space can be calculated from the angular power spectrum. We describe their relations following \citet{sanchezcid2026darkenergysurveyyear},
\begin{align}
w^{ij}(\theta) &=
\sum_{\ell} \frac{2\ell+1}{4\pi}\,
P_{\ell}(\cos\theta)\,
C^{ij}_{\delta_{g,\mathrm{obs}},\delta_{g,\mathrm{obs}}}(\ell),\nonumber
\\
\gamma_t^{ij}(\theta) &=\sum_{\ell}
\frac{2\ell+1}{4\pi}\,
\frac{P_{\ell}^{2}(\cos\theta)}{\ell(\ell+1)}\,
C^{ij}_{\delta_{g,\mathrm{obs}},E}(\ell),\nonumber
\\
\xi_{\pm}^{ij}(\theta) &=\sum_{\ell}\frac{2\ell+1}{4\pi}\,
\frac{2\bigl(G^{+}_{\ell,2}(x)\pm G^{-}_{\ell,2}(x)\bigr)}
{\ell^2(\ell+1)^2}\,\nonumber\\
&\cdot\bigl[C^{ij}_{EE}(\ell)\pm C^{ij}_{BB}(\ell)\bigr],
\label{eq:fourier2real}
\end{align}
where $P_\ell$ are the Legendre polynomials and $G_\ell$ are the associated Legendre functions/polynomials. $\mathrm{E}$ and $\mathrm{B}$ refer to the E- and B-modes of the shear field, defined as
\begin{align}
    \gamma_{\mathrm{E}/\mathrm{B}}^i(\boldsymbol{\ell}) = (\delta_{\alpha\beta},\epsilon_{\alpha\beta})T_\alpha(\boldsymbol{\ell})\gamma_\beta^i(\boldsymbol{\ell}),
\end{align}
where $\boldsymbol{\mathrm{T}}(\boldsymbol{\ell})\equiv (\cos{(2\phi_\ell)},\sin{(2\phi_\ell)})$, $\phi_\ell$ is the angle of the $\boldsymbol{\ell}$ vector from the $\ell_x$ axis, $\delta_{\alpha\beta}$ is the Kronecker delta function, and $\epsilon_{\alpha\beta}$ is the two-dimensional Levi-Civita tensor.

It is impossible to measure the correlation functions at an exact angular separation because of finite galaxy samples, so in practice they are estimated over angular bins, such as $[\theta_\mathrm{min},\theta_\mathrm{max}]$. To match that on the theoretical side, the predicted correlation functions need to be averaged over the same angular bins. We denote this as the bin-average correction, which replaces the Legendre functions in Equation~\ref{eq:fourier2real} with their bin-averaged versions $\overline{P_\ell}$, $\overline{P_\ell^2}$, and $\overline{G_{\ell,2}^+ \pm G_{\ell,2}^-}$, e.g.,
\begin{align}
    \overline{P_\ell}(\theta_\mathrm{min},\theta_\mathrm{max})&\equiv\frac{\int_{\cos{\theta_\mathrm{min}}}^{\cos{\theta_\mathrm{max}}} dx P_\ell(x)}{\cos{\theta_\mathrm{max}}-\cos{\theta_\mathrm{min}}}\nonumber\\
    &= \frac{[P_{\ell+1}(x) - P_{\ell-1}(x)]|_{\cos{\theta_\mathrm{min}}}^{\cos{\theta_\mathrm{max}}}}{(2\ell+1)(\cos{\theta_\mathrm{max}} - \cos{\theta_\mathrm{min}})}.
\end{align}

For small sky patches, it is computationally efficient to replace brute-force sums over the Legendre functions with integrals of the Bessel functions of the first kind without losing too much precision. This is known as the flat-sky approximation.

\begin{align}
    w^{ij}(\theta) &= \int_0^\infty \frac{d\ell \ell}{2\pi} C_{\delta_{g,\mathrm{obs}},\delta_{g,\mathrm{obs}}}^{ij} J_0(\ell\theta), \\
    \gamma_t^{ij}(\theta) &= \int_0^\infty \frac{d\ell \ell}{2\pi} C_{\delta_{g,\mathrm{obs}},E}^{ij} J_2(\ell\theta), \\
    \xi_{\pm}^{ij}(\theta) &= \int_0^\infty \frac{d\ell \ell}{2\pi}C_{EE}^{ij}J_{0/4}(\ell\theta),
\end{align}
where $J_n$ are the Bessel functions of the first kind.

In addition to the basic modeling of the tracers, we also include the following astrophysical and observational effects. Note that we do not study these components, although they are also approximate models, similar to those discussed in Section~\ref{sec:approx}.
\begin{itemize}

\item \textbf{Intrinsic alignment (IA)}

The gravitational shear field $\gamma_{\alpha,\mathrm{G}}$ is correlated with the galaxy intrinsic shear field, also known as the intrinsic ellipticity $\gamma_{\alpha,\mathrm{I}}$. The latter thus has a non-negligible contribution to the cosmic shear signal, which should be taken into account in the observed weak lensing shear field in Equation~\ref{gravitational-shear-field-1} and~\ref{gravitational-shear-field-2}.
\begin{align}\label{weak-lensing-shear-field-1}
    \gamma_{\alpha,\mathrm{obs}} = \gamma_{\alpha,\mathrm{G}} + \gamma_{\alpha,\mathrm{I}}.
\end{align}

We parameterize and marginalize out this effect using the nonlinear linear alignment model \citep[NLA,][]{HirataSeljak2004, BridleKing2007}, which assumes that the three-dimensional intrinsic shear field $\tilde{\gamma}_{\alpha,\mathrm{I}}$ is proportional to the linear term of the gravitational tidal field $s_\alpha$. 

We can therefore model the three-dimensional intrinsic shear field as, 
\begin{align}
    \tilde{\gamma}_{\alpha,\mathrm{I}} = -a_1\bar{C}_1\frac{\rho_{\mathrm{crit}}\Omega_\mathrm{m}}{D(z)}(\frac{1+z}{1+z_0})^{\eta_1} s_\alpha,
\end{align}
where the pivot redshift $z_0$ is the mean redshift of the source-galaxy sample, $\rho_\mathrm{crit}$ is the critical density, $\bar{C}_1 = 5\times10^{-14}\,M_\odot h^{-2} \mathrm{Mpc}^2$ \citep{Secco_2022} is a normalization constant, and $D(z)$ is the linear growth factor. We marginalize over two nuisance parameters in this model: the dimensionless amplitude $a_1$ and the power-law index $\eta_1$.

DES-Y6 adopts the TATT model \citep{PhysRevD.100.103506}, which includes contributions from the tidal-torquing term and validates that the cosmological constraints are robust to the choice between NLA and TATT \citep{DESY63x2pt2026}. We therefore adopt the NLA model and argue it is adequate for the purposes of our analysis.

\item \textbf{Redshift and shear calibration}

We employ a constant shift $\Delta z$ in each tomographic bin of the redshift distribution $n(z)$ to describe its uncertainty,
\begin{align}
    n(z)\rightarrow n(z-\Delta z).
\end{align}
We marginalize over one shift parameter per tomographic bin for both lens and source samples.

In weak lensing shape measurements, the bias introduced by a range of observational effects, such as PSF modeling, blending, and the under-sampling of the image, can be modeled as a multiplicative bias, as
\begin{align}
    \gamma_{\alpha,\mathrm{obs}} = (1+m)(\gamma_{\alpha,\mathrm{G}} +\gamma_{\alpha,\mathrm{I}}).
\end{align}
We marginalize over one shear calibration bias parameter per tomographic bin of the source-galaxy sample.

\item \textbf{Linear galaxy bias}

We assume a linear relation between the galaxy overdensity field and the matter overdensity field,
\begin{align}
    \delta_g = b_g\delta_m.
\end{align}
We assume one galaxy bias parameter per tomographic bin of the lens-galaxy sample. The nonlinear galaxy bias model, motivated by perturbation theory \citep{Pandey_2020}, provides more information on small scales. However, DES-Y6 finds that the results of linear and nonlinear galaxy bias are consistent, and the latter does not significantly increase the constraining power \citep{sanchezcid2026darkenergysurveyyear}. We therefore choose linear galaxy bias for simplicity and do not expect this choice to affect our conclusions.

\end{itemize}

\subsection{Approximations}
\label{sec:approx}
\vspace{0.2cm}
\subsubsection{Limber approximation}
\vspace{0.2cm}
The calculation of the angular power spectrum involves integrals over the spherical Bessel functions. For example, the integrand in Equation~\ref{angular-power-spectrum-gg} is highly oscillatory, which makes numerical integration challenging. However, in the limit of $x>\ell \gg 1$, $j_\ell(x)$ contributes mostly at $x=\ell+\frac{1}{2}$. Therefore, the spherical Bessel function can be approximated as a Dirac delta function,
\begin{align}
    j_\ell(k\chi) \simeq \sqrt{\frac{\pi}{2\ell+1}} \delta_D(k\chi-\ell-\frac{1}{2}).
\end{align}
When this approximation is applied to the product of two spherical Bessel functions, we use the following identity,
\begin{align}
    \delta_D(k\chi_1-\ell-\frac{1}{2})\delta_D&(k\chi_2-\ell-\frac{1}{2}) \nonumber\\
  &= \frac{\delta_D(k-(\ell+1/2)/\chi_1)}{\chi_1}\frac{\delta_D(\chi_1-\chi_2)}{k}.
    \label{limber-integration}
\end{align}
The angular power spectrum thus has a general form under the Limber approximation,
\begin{align}
    C_{AB}^{ij}(\ell) = \int d\chi\frac{W_A^i(\chi)W_B^j(\chi)}{\chi^2} P_{AB}(k=\frac{\ell+1/2}{\chi}, z(\chi)),
    \label{eq:cl-limber}
\end{align}
where $A$ and $B$ denote the observed fields, and $P_{AB}$ is the corresponding three-dimensional power spectrum.

However, there are a few physical limitations to this approximation: (1) it is applicable only when $\ell\gg 1$. (2) it requires that the weight function $W(\chi)$ is wide enough so that the integration can cover multipoles much greater than 1, i.e. $k\Delta\chi\gg1$. Therefore, the assumption that the oscillatory parts of $j_\ell(x)$ cancel each other and that only the parts on $x\sim \ell+\frac{1}{2}$ mostly effectively contribute to the integral. (3) the weight functions between two observables should largely overlap.

For cosmic shear and galaxy-galaxy lensing, the Limber approximation has been shown to be sufficiently accurate for surveys such as Rubin LSST-Y1 and DES-Y6 \citep{Fang_2020}, since their galaxy-sample distributions are redshift-wide and their statistical power is much lower than that of the final-year Stage-IV surveys. However, neither of the above approximations has been shown to hold for Roman HLIS-Y5 and Rubin LSST-Y10. We thus only implement the non-Limber algorithm for galaxy clustering in this work, which we justify in Section~\ref{sec:code_comparison}. The Non-Limber algorithm we used is supported by the Fast Fourier Transform in Logarithmic space (FFTLog) algorithm \citep{Fang_2020}. Like Equation~\ref{angular-power-spectrum-gg}, we need to solve an integral in a general form,
\begin{align}
    \int &d\chi_1 W^A(\chi_1) \int d\chi_2 W^B(\chi_2)\nonumber\\
    &\cdot\int \frac{2k^2}{\pi}dk P^{AB}(k,z_1,z_2)j^{(n^A)}_\ell(k\chi(z_1))j^{(n^B)}_\ell(k\chi(z_2)),\nonumber\\
\end{align}
where the $P^{AB}$ is the power spectrum of the observed fields $A$ and $B$, and it is related to the nonlinear matter power spectrum. 

The usual approach is to divide $P^{AB}$ into two parts, a linear part, which depends on the linear matter power spectrum $P_{\mathrm{lin}}$, and a nonlinear part, which depends on the nonlinear contribution from the matter power spectrum $P_{\delta} - P_{\mathrm{lin}}$. The nonlinear part can be directly computed by implementing the Limber approximation because it works well on small scales, where the nonlinear contribution is significant. For the linear part, we can decompose it into a redshift-independent component and redshift-dependent transfer functions. This decomposition allows us to rewrite the double-Bessel integral in Equation~\ref{angular-power-spectrum-gg} to two one-dimensional single Bessel integrals, which can be computed efficiently using $\mathrm{FFTLog}$.

 \begin{figure*}
  \centering
  \includegraphics[width=\textwidth]{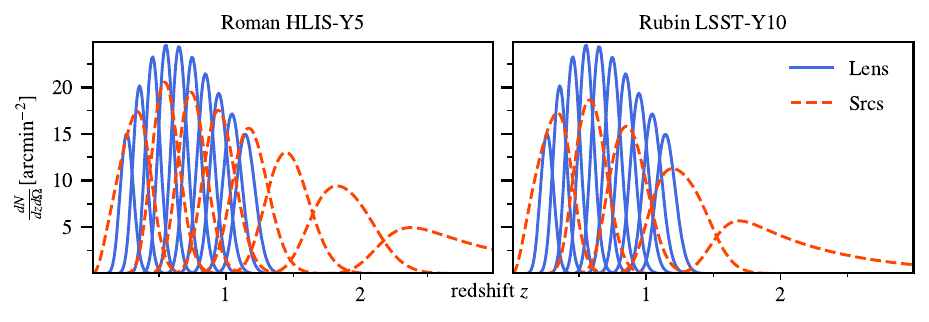}
  \caption{These plots show the redshift distributions of galaxy samples for Roman (left) and Rubin (right) used in this work, as defined in Section~\ref{sec:roman} and \ref{sec:rubin}.
  }
  \label{fig:nofz}
\end{figure*} 

\subsubsection{Redshift-space distortions}
\label{sec: redshift-space-distortion}
\vspace{0.2cm}
The second effect we consider is a correction to the projected galaxy overdensity field due to the galaxy peculiar velocity with respect to the Hubble flow, an effect referred to as redshift-space distortions (RSD). We follow the formalism of \citet{Fisher_1994} and \citet{Padmanabhan_2007}. Consider a galaxy located at a comoving distance $\chi$ with a line-of-sight peculiar velocity $v_{\parallel} (\chi)= \boldsymbol{v}(\chi)\cdot\boldsymbol{\hat{n}}$ relative to the Hubble flow. In the non-relativistic limit, $v_\parallel\ll c$, this velocity introduces an additional redshift $1+z_{\mathrm{pec}} =1+v_\parallel(\chi)/c$. Combining this with the cosmological redshift, $1+z_{\mathrm{hub}} \equiv 1/a(\chi)$, gives the observed redshift of the galaxy,
\begin{align}
    z_\mathrm{obs}&=(1+z_\mathrm{pec})(1+z_\mathrm{hub})-1 \nonumber\\
                  &= z_\mathrm{hub} + v_\parallel(\chi)/(a(\chi)c).
\end{align}
Therefore, the observed galaxy redshift is shifted by,
\begin{align}
    \delta z=v_\parallel/(a(\chi)c).
\end{align}
When galaxies are selected by the observed redshift, this shift can be represented as a correction to the selection function,
\begin{align}
    W_{\delta,g}(\chi)\rightarrow W_{\delta,g}(\chi+\delta \chi),\qquad\delta\chi \approx c\delta z/H(\chi).
\end{align}
With this correction, Equation~\ref{galaxy-overdensity-field} becomes,
\begin{align}
    \delta_{g,\mathrm{obs}}^i &= \int d\chi W_{\delta,g}^i(\chi+\delta\chi) (1+\delta_g^{(3D)}(\boldsymbol{\hat{n}}\chi,\chi)) - 1\nonumber\\
    &= \int d\chi (W_{\delta,g}^i + \frac{\partial W_{\delta,g}^i}{\partial\chi}\delta\chi)(1+\delta_g^{(3D)}(\boldsymbol{\hat{n}}\chi,\chi)) - 1\nonumber\\
    &\approx \delta_{g,D}^i(\boldsymbol{\hat{n}}) - \int d\chi W_{\delta,g}^i \frac{\partial}{\partial\chi}(\frac{\boldsymbol{\hat{n}}\cdot\boldsymbol{v}(\boldsymbol{\hat{n}}\chi,\chi)}{a(\chi)H(\chi)})\nonumber\\
    &= \delta_{g,D}^i + \delta_{g,\mathrm{RSD}}^i. \label{galaxy-overdensity-field-RSD}
\end{align}
In the second line of Equation~\ref{galaxy-overdensity-field-RSD}, we regard $\delta\chi$ as a small perturbation and expand the normalized selection function using the Taylor expansion. We denote the second term of the last line of Equation~\ref{galaxy-overdensity-field-RSD} as $\delta_{g,\mathrm{RSD}}$, accounting for the linear contribution from RSD to the galaxy overdensity field.

The derivative in the third line of Equation~\ref{galaxy-overdensity-field-RSD} can be approximated as $f\mu^2\delta_m$ in harmonic space under a linear approximation, where $\mu$ is defined as $k_\parallel/k$ and $f$ is the linear growth rate.

\subsubsection{Nonlinear matter power spectrum}
\vspace{0.2cm}
An accurate nonlinear matter power spectrum is critical for recovering the true cosmology in the 3$\times$2pt inference pipeline, as it describes the matter distribution on smaller scales. 
In most state-of-the-art 3$\times$2pt analyses \citep{DESY63x2pt2026, Heymans2021_KiDS1000_3x2pt}, \textsc{HMCode} \citep{Mead_2021} has been the model used. However, the recent development of \textsc{EuclidEmulator2} \citep{EuclidEmulator2_2021} has shown that better, higher-resolution N-body simulations can result in more accurate predictions for the nonlinear matter power spectrum. As such, we use \textsc{EuclidEmulator2} as our baseline nonlinear matter power spectrum model and test the impact of using \textsc{HMCode} as an alternative model. 

Below, we briefly introduce these two models.

\begin{itemize}
\item \textbf{\textsc{EuclidEmulator2}}

\textsc{EuclidEmulator2} provides precise predictions of the nonlinear matter power spectrum within an eight-dimensional cosmological parameter space, including: $\Lambda\mathrm{CDM}$ cosmological parameters $(\Omega_\mathrm{b},\Omega_\mathrm{m},A_s,n_s,h)$, dark energy equation-of-state parameters $(w_0,w_a)$, and the sum of neutrino masses $\sum m_\nu$. \textsc{EuclidEmulator2} is a regression model, which is trained by optimizing the coefficients of a polynomial chaos expansion (PCE) to predict the weights of 14 PCA bases of the matter power spectrum from cosmological parameters. The actual training is done with a dataset including 127 high-resolution simulations based on a pre-trained model, the training dataset of which is from \textsc{HaloFit} \citep{Takahashi2012}.

The \textsc{EuclidEmulator2} predictions achieve 1\% accuracy for the nonlinear correction factors for $z\leq3$ and $0.01\,h \mathrm{Mpc}^{-1}\leq k\leq10\,h\mathrm{Mpc}^{-1}$, and it is so far the most precise nonlinear matter power spectrum model among those widely used by the weak lensing community. Since \textsc{EuclidEmulator2} has a strict boundary on $k$-modes, we calculate the power spectrum of $k$-modes exceeding 10 $h\mathrm{Mpc}^{-1}$ with \textsc{HMCode}. Since \textsc{EuclidEmulator2} has a strict boundary of cosmological parameters, we follow that to set the the parameter priors as shown in Table~\ref{tab:parameters}.

\item \textbf{\textsc{HMCode}}

We choose \textsc{HMCode} as an alternative model of the nonlinear matter power spectrum, as it provides analytic and precise reconstructions, with an RMS error less than 2.5 percent over a wide range of cosmologies for scales $k<10\,h\mathrm{Mpc}^{-1}$ and redshifts $z<2$. \textsc{HMCode} is reliable in non-standard cosmologies since it predicts the nonlinear matter power spectrum analytically. It also provides a parameterization prescription to model the impact of baryonic feedback. However, we do not use this feature, as our scale cuts should eliminate the majority of the contribution from baryons.

\textsc{HMCode} is based on the halo model \citep{Seljak_2000} and modifies it with a series of parameterizations to match the actual power spectrum from simulations. The parameters include those controlling the mass function, halo concentration, properties of the linear power spectrum, and a spectral index that smooths the transition between the one- and two-halo terms, etc. The parameterization is independent of cosmological parameters, which enables a maximum applicability to novel cosmological scenarios, though some parameters related to the nonlinear matter power spectrum are calibrated by 37 representative simulations.

\end{itemize}
 \vspace{0.15cm}
\section{Survey specification}
\label{sec:surveys}
\vspace{0.2cm}
In this section we briefly describe the baseline design and data characteristics used in our forecasting exercise for Roman (Section~\ref{sec:roman}) and Rubin (Section~\ref{sec:rubin}). The goal is to model the source and lens samples for a ``fiducial'' 3$\times$2pt analysis for the \textit{final} Roman and Rubin dataset, assuming 5 years for Roman HLIS data (Roman HLIS-Y5) and 10 years for Rubin data (Rubin LSST-Y10).  

For Rubin we mostly follow the LSST Dark Energy Science Collaboration (DESC) Science Requirements Document (SRD) \citep{thelsstdarkenergysciencecollaboration2021lsstdarkenergyscience}. For Roman, there have been a number of proposed source and lens samples in the literature with different analysis choices \citep{Eifler_2021,Cao2026}. Without loss of generality, here we opt to use the source sample defined in the Data Challenge 1\footnote{The DC1 is to study the effect of model mis-specification by estimating how the inference is biased by the data vectors which are computed beyond or at the extreme values of the model's parameter space. It tests a wide range of model components including baryonic feedback, intrinsic alignment, galaxy bias, shear calibration, and photo-z uncertainty. It also studies the effect of prior choices and scale cuts.} (DC1, manuscript in preparation) of the Cosmological Parameters Inference Pipeline (CPIP) working group (WG) of Roman HLIS Cosmology PIT and the same lens sample used in the Rubin case. We show in Appendix~\ref{app:analysis_choice} the impact of using the different lens samples.

The final redshift distributions used for the two surveys in this paper are shown in Figure~\ref{fig:nofz} and the survey specifications are summarized in Table~\ref{tab:survey_settings}. 

\begin{table}
    \centering
    \caption{The survey settings for Roman HLIS-Y5 and Rubin LSST-Y10. For Roman HLIS-Y5, its survey area corresponds to the medium tier. We assume both surveys use the Rubin LSST-Y10 lens sample, the number density of which decreases from 48 to 30 $\mathrm{arcmin}^{-2}$ after the tomographic binning.
    }
    \label{tab:survey_settings}
    \begin{tabular}{|ccc|}
    \hline
         & Roman HLIS-Y5 & Rubin LSST-Y10\\
         \hline
    Survey area ($\mathrm{deg}^2$) & 2,415 & 14,300 \\
    Lens number density $n_{\mathrm{eff}}$  & 48 (30) & 48 (30) \\
    Source number density $n_{\mathrm{eff}}$  & 41.00 & 27.73 \\
    Shape noise $\sigma_e$ & 0.26 & 0.26 \\
    Number of lens bins & 10 & 10 \\
    Number of source bins & 8 & 5 \\
    Number of $\theta$ bins & 20 & 20 \\
    $\theta$ range ($\mathrm{arcmin}$)& $(1, 500)$ & $(1, 500)$ \\
    Number of $\ell$ bins & 20 & 20 \\
    $\ell$ range & (20, 4000) & (20, 4000) \\
    \hline
    \end{tabular}
\end{table}

\subsection{Roman}
\label{sec:roman}
\vspace{0.2cm}
The 3$\times$2pt analysis in Roman will use the imaging data from Roman's High Latitude Wide Area Survey (HLWAS), referred to as High Latitude Imaging Survey (HLIS). In the most recent design \citep{ROTAC_2025_FinalReport}, HLIS will include four survey tiers: medium, deep, ultra-deep and wide. We focus on the medium tier, with an effective source-galaxy number density of $n^{s}_{\mathrm{eff}}\approx 41\,\mathrm{arcmin}^{-2}$, which covers $2,415 \;\mathrm{deg}^2$ of sky and reaches a characteristic imaging depth of $m_\mathrm{AB} = 26.5$ in three imaging bands (Y,J,H) for a $5\sigma$ point-source. 

Following DC1, we generate the source-galaxy redshift distribution with the Galaxy Survey Exposure Time Calculator\footnote{\url{https://roman.gsfc.nasa.gov/science/etc14.html}} (ETC). This ETC is first applied to the CANDELS catalog \citep{Guo_2013} with the same S/N cuts as \citet{Eifler_2021} to get the distribution of detected galaxy samples. We then fit its shape, split it into 8 redshift bins with an equivalent number density per bin, and convolve each bin with a Gaussian distribution with a width of 0.05.

\subsection{Rubin}
\label{sec:rubin}
\vspace{0.2cm}
The 3$\times$2pt analysis for Rubin will use the Wide-Fast-Deep (WFD) component from the Legacy Survey of Space and Time (LSST). LSST is estimated to achieve $5\sigma$ point-source detection depths of 25.30, 26.84, 27.04, 26.35, 25.22, 24.47 in ($u,g,r,i,z,y$) over $14,300\;\mathrm{deg}^2$ of sky. It is expected to provide an effective number density of $n_{\mathrm{eff}}^{l} \approx 48\;\mathrm{arcmin}^{-2}$ for lens-galaxy samples and $n_{\mathrm{eff}}^{s} \approx 27\;\mathrm{arcmin}^{-2}$ for source-galaxy samples. To describe the redshift distributions for the galaxy samples, we use the analytical Smail model \citep{Smail_1994} to specify the distributions, as done in the LSST DESC SRD \citep{thelsstdarkenergysciencecollaboration2021lsstdarkenergyscience},
\begin{equation}
    \frac{dN}{dz} \propto z^2\mathrm{exp}{[-(z/z_0)^\alpha]}.
\label{sec:nofz}
\end{equation}

For the lens sample, we set $z_0=0.28$ and $\alpha=0.90$ in Equation~\ref{sec:nofz}, and require the total number density to equal $48\;\mathrm{arcmin}^{-2}$. The fitted redshift distribution is then uniformly divided into 10 tomographic bins between redshift $0.2\leq z\leq1.2$, therefore, some samples at the lowest and highest redshifts are lost and the rest effective number density is about $30\;\mathrm{arcmin}^{-2}$.

For the source sample, we set $z_0=0.11$ and $\alpha=0.68$, and require the total number density to equal $27\;\mathrm{arcmin}^{-2}$. The fitted redshift distribution is then divided into 5 tomographic bins with equivalent number density per bin.

\section{Simulated likelihood analysis}
\label{sec:likelihood}
\vspace{0.2cm}
In this section we describe our simulated likelihood analysis. We primarily use the \textsc{Cobaya}-\textsc{CosmoLike} Architecture\footnote{\url{https://github.com/CosmoLike/cocoa}} (\textsc{CoCoA}) for this purpose. \textsc{CoCoA} is built upon the cosmology library \textsc{CosmoLike} \citep{Krause2017} to simulate the 3$\times$2pt data vectors and employs the COde for BAYesian Analysis framework \citep[\textsc{Cobaya},][]{Torrado2021} to perform the inference. Besides the default MCMC sampler of \textsc{Cobaya}, \textsc{CoCoA} also integrates a series of sampler tools, including \textsc{PolyChord}, \textsc{Nautilus} and \textsc{Emcee}.

It is worth noting that \textsc{CoCoA} also integrates a wide range of analysis tools which are intensively used in the cosmological inference, such as likelihood maximizer, likelihood profile, and tension metrics. it will also provide a set of emulators for commonly used inference likelihoods and the \textsc{CAMB}\footnote{\url{https://github.com/cmbant/CAMB}} to accelerate the sampling process. One of the main features of \textsc{CoCoA} is that it introduces shell scripts that allow users to containerize \textsc{Cobaya}, the Boltzmann codes, and multiple likelihoods, ensuring that users can adopt consistent versions for the Fortran/C/C++ compilers and libraries across multiple machines.

We use the Core Cosmology Library\footnote{\url{https://github.com/LSSTDESC/CCL}} \citep[\textsc{CCL,}][]{Chisari_2019} as a reference pipeline. It is a public standardized library of routines to calculate the basic observables, which will be the standard theoretical prediction package for LSST DESC. In Section~\ref{sec:code_comparison}, we compare our model and \textsc{CCL} at the data vector level, ensuring that no major modeling mismatch exists. 

In this work we perform simulated likelihood analysis for Roman HLIS-Y5 and Rubin LSST-Y10 with configurations as described in Section~\ref{sec:surveys}. We examine how the $\Lambda$CDM constraints change when we employ the Limber approximation, omitting RSD, and change the nonlinear matter power spectrum model. We also examine these results in both real and harmonic space. To perform this study, we first describe in Section~\ref{sec:scale_cut} our procedures for determining the scale cuts for the different 3$\times$2pt data vectors. We next describe in Section~\ref{sec:inference} our inference framework and what we choose to use for the various nuisance parameters.  

\subsection{scale cuts}
\label{sec:scale_cut}
\vspace{0.2cm}
\begin{table}
    \centering
    \caption{The signal-to-noise ratio (S/N) or the $\sqrt{\Delta\chi^2}$ for each probe after our fiducial scale cuts. The $\Delta\chi^2$ is defined as $\Delta\chi^2 = \mathbf{M(p)}^T\mathbf{Cov}^{-1}\mathbf{M(p)}$. The difference between real- and harmonic-space S/N values is mainly due to the scale cuts.}
    \begin{tabular}{|lcc|}
        \hline
        Probe & Roman HLIS-Y5 & Rubin LSST-Y10 \\
        \hline
        $\xi^\pm$      & 72 & 101  \\
        $\gamma_t$   & 157 & 340 \\
        $w(\theta)$  & 122 & 297\\
        $\mathrm{total}$ & 191 & 427 \\
        \hline
        $C_{\ell}^{ss}$  & 46  & 64 \\
        $C_{\ell}^{gs}$  & 180 & 389 \\
        $C_{\ell}^{gg}$  & 302 & 738 \\
        $\mathrm{total}$ & 332 & 791 \\
        \hline
    \end{tabular}
    \label{tab:lsst_roman_s/n}
\end{table}

Unmodeled systematics on small scales are one of the main sources of theoretical uncertainties. For cosmic shear, the main uncertainty comes from our (lack of) understanding of the baryon distribution on small scales due to baryonic feedback \citep{Chisari2018, RS2025}. For two-point functions involving the galaxy density, the main uncertainty comes from the nonlinear galaxy bias.  

There are a wide range of approaches in the literature that attempt to model these small-scale information \citep{Schneider2015, Arico2021}. In general, although recent results have shown promise in constraining these small-scale effects with external data \citep{xu2026, amy2025}, the community as a whole has not converged on an agreed-upon model and a set of agreed-upon priors. As such, in this analysis we use a slightly conservative approach that is well-validated by the Stage-III surveys -- we remove the small scales that are uncertain in our analysis.   

We note that there have been previous attempts to perform Roman and Rubin forecasts that have adopted different approaches towards scale cuts. For example, the DESC SRD has assumed $\ell_{\mathrm{max}}=3000$ for cosmic shear and $k_{\rm max}=0.3\, h/{\rm Mpc}$, or equivalently, $R_{\rm min}=21 \,\mathrm{Mpc}/h$ for galaxy-galaxy lensing and galaxy clustering. In \citet{Eifler_2021, Cao2026}, they used the same assumptions as the DESC SRD for Roman forecasts. In this work, we use an approach more similar to what is done in the DES-Y6 cosmology analysis \citep{DESY63x2pt2026}. We compare the constraining power of different scale cuts in Appendix~\ref{app:analysis_choice}.  

We now describe how we define scale cuts in real and harmonic spaces, and how we convert between the two.

For cosmic shear, we set a $\Delta\chi^2$ budget for each tomographic bin pair and thus iteratively determine the scale cuts by requiring the $\Delta\chi^2$ between the contaminated and fiducial data vectors to fall below the budget. Specifically, we generate baryon-contaminated cosmic shear from \textsc{Illustris} simulations \citep{Nelson_2015} and set $\Delta\chi^2=0.025$ budget per bin pair in real space, while we use 0.05 in harmonic space to balance the total $\Delta\chi^2$ in both spaces. 

For galaxy-galaxy lensing and galaxy clustering, we set a physical minimum transverse radius $(R_\mathrm{min}^{\gamma t}, R_\mathrm{min}^{w}) = (6 \,{\rm Mpc}/h, 8 \,{\rm Mpc}/h)$. We take the approximation $\theta_\mathrm{min} = R_\mathrm{min}/ \chi(\bar{z})$ to convert the minimum transverse radius into the minimum angle in real space, where $\bar{z}$ is the effective average redshift of a lens-galaxy bin. We then utilize the relation $\ell_\mathrm{max} = \pi/\theta_\mathrm{min}$ to convert the minimum angle in real space to the largest multipole in harmonic space, i.e., the scale cuts in harmonic space. For both Roman and Rubin, we measure 20 angular bins in the angular range $1'<\theta < 500'$ in real space while measuring 20 multipoles in the range $20<\ell<4000$ in harmonic space. Table~\ref{tab:lsst_roman_s/n} shows the signal-to-noise (S/N) values for each part of the data vectors after the scale cuts.

\begin{table}
    \centering
    \caption{Model parameters for the analysis. Superscript $i$ refers to the index of the tomographic bin. $\mathcal{U}(a,b)$ represents a uniform distribution between $a$ and $b$, while $\mathcal{N}(\mu,\sigma^2)$ represents a Gaussian distribution centered at $\mu$ with a variance $\sigma^2$.}
    \begin{tabular}{|lcc|}
    \hline
     Parameter & Fiducial & Prior \\
     \hline
     $A_s\times10^9$ & 2.1 & $\mathcal{U}(1.7,2.5)$\\
     $n_s$ & 0.96605 & $\mathcal{U}(0.92,1.00)$\\
     $H_0$ & 67.32 & $\mathcal{U}(61,73)$\\
     $\Omega_\mathrm{b}$ & 0.0495 & $\mathcal{U}(0.04,0.06)$\\
     $\Omega_\mathrm{m}$ & 0.316 & $\mathcal{U}(0.24,0.40)$\\
    Lens photo-z $\Delta z^i$ & $\{0\}^i$ & $\mathcal{N}(0, 0.003^2)$\\
    Source photo-z $\Delta z^i$ & $\{0\}^i$ & $\mathcal{N}(0, 0.001^2)$\\
    Source calibration $m^i$ & $\{0\}^i$ & $\mathcal{N}(0, 0.003^2)$\\
    Intrinsic Alignment $a_\mathrm{1},\eta_1$
     & 0.5, 0 & $\mathcal{U}(-5,5)$,$\mathcal{U}(-5,5)$ \\
    Galaxy Bias $b^i$ & $\frac{0.95}{D(\bar{z}^i)}$ & $\mathcal{U}(0.8, 2)$ \\
    \hline
    \end{tabular}
    \label{tab:parameters}
\end{table} 

\vspace{0.15cm}
\subsection{Covariance matrix and inference}
\label{sec:inference}
\vspace{0.2cm}
We use \textsc{CoCoA} to perform the inference of parameters $\mathbf{p}$ from the observed 3$\times$2pt data $\mathbf{D}$. Throughout we assume a Gaussian likelihood of the form, 
\begin{align}
  \mathcal{L}(\mathbf{D}|\mathbf{p}) \propto \exp\left(-\frac{1}{2}\left[\mathbf{D} - \mathbf{M}(\mathbf{p})\right]^\top \mathbf{Cov}^{-1} \left[\mathbf{D} - \mathbf{M}(\mathbf{p})\right]\right),
  \label{eq:likeli}
\end{align}
where $\mathbf{M}(\mathbf{p})$ is the model for the observables, and $\mathbf{Cov}$ is the covariance matrix for the data.  We now describe the model, its parameters, their priors, and the covariance matrix, as well as the tools used for the inference.

\begin{figure*}
  \centering
  \includegraphics[width=1.0\linewidth]{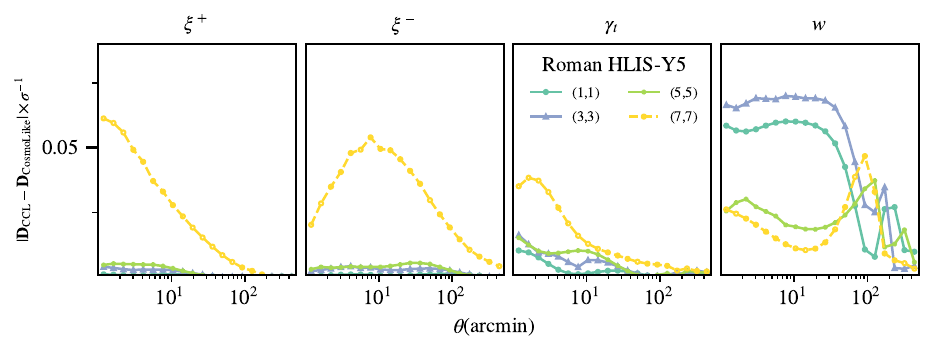}
  \caption{This figure shows the bias significance of data vector between two pipelines over the angular scale: \textsc{CCL} and \textsc{CoCoA}. The data vector difference is first computed and then divided by the square root of diagonal term of data vector covariance matrix to get the significance of bias. Each of the four subplots represents a probe, including $(\xi^+, \xi^-, \gamma_t, w)$. Only part of the correlations are shown in the figure, for example, (1,1) represents the auto-correlation of the first source bin for cosmic shear, or the correlation between the first lens bin and the first source bin for galaxy-galaxy lensing, or the auto-correlation of the first lens bin for galaxy clustering.}
  \label{fig:Code-comparison-roman-real}
\end{figure*}

\begin{itemize}
    \item \textbf{Model}

    $\mathbf{M}(\mathbf{p})$ comprises three power spectra of probes, cosmic shear $C^{ss}$, galaxy-galaxy lensing $C^{gs}$, and galaxy clustering $C^{gg}$ as described in Section~\ref{sec:model}. Then they are converted to the correlation functions $(\xi^+,\xi^-,\gamma_t,w(\theta))$ for real-space analyses by Equation~\ref{eq:fourier2real}.
    
    \item \textbf{Parameters and priors}

    To simulate the data vector $\mathbf{M}(\mathbf{p})$, cosmological and nuisance parameters $\mathbf{p}$ are required. In this paper, we sample and marginalize over five $\Lambda \mathrm{CDM}$ cosmological parameters ($A_s$, $n_s$, $H_0$, $\Omega_\mathrm{b}$, $\Omega_\mathrm{m}$) as described in Section~\ref{sec:model}, while fixing other cosmological parameters (the sum of neutrino masses $m_\nu$, the equation-of-state parameter of dark energy $w$, and the reionization optical depth $\tau$). 

    For the astrophysical nuisance parameters, we set broad, flat priors since there are still large uncertainties in our understanding of these effects. We set priors of the amplitude and power index parameters of the NLA intrinsic alignment model to be $\mathcal{U}(-5,5)$, and priors on the linear galaxy bias parameters to be $\mathcal{U}(0.8, 2)$. 

    For observational nuisance parameters associated with redshift and shear calibration, we set strong Gaussian priors given advances in Stage-III surveys -- most of these effects now have comparable small uncertainties. We use Gaussian priors $\mathcal{N}(\sigma=0.003^2)$ for the lens-galaxy redshift uncertainty and $\mathcal{N}(\sigma=0.001^2)$ for the source-galaxy redshift uncertainty. We use $\mathcal{N}(\sigma=0.003^2)$ for shear calibration uncertainty. These numbers are based on the DESC SRD but are assumed to be similar for Roman. 
    
    The fiducial values and priors of these parameters are described in Table~\ref{tab:parameters}.
    
    \item \textbf{Covariance matrix}

    We analytically simulate the covariance matrix using the package \textsc{CosmoCov} \citep{Krause2017, Fang__2020, Fang_2020}. It decomposes the covariance matrix into three parts, the Gaussian covariance, non-Gaussian covariance, and the super-sample covariance as
    \begin{equation}\label{eq:covariance}
    \begin{aligned}
        {\ \ \ \ }&\mathrm{Cov}(C(\ell_1),C(\ell_2)) = \mathrm{Cov}^\mathrm{G}(C(\ell_1),C(\ell_2)) \nonumber\\&+ \mathrm{Cov}^\mathrm{NG,0}(C(\ell_1),C(\ell_2))+
        \mathrm{Cov}^\mathrm{SSC}(C(\ell_1),C(\ell_2)).
    \end{aligned}
    \end{equation}
    The equations for calculating the covariance of the three probes, cosmic shear, galaxy-galaxy lensing, and galaxy clustering, can be found in \citet{krause2017darkenergysurveyyear}. The transformation of covariances from harmonic space to real space follows \citet{Fang_2020}, which employs the $\mathrm{2D-FFTLog}$ algorithm to accelerate the double Bessel integration involved in this process.
    
    \item \textbf{Sampling algorithm}

    We employ the Metropolis sampler used by \textsc{CosmoMC} \citep{Lewis:2013hha} throughout this paper. We assess the convergence of the chains by Gelman-Rubin statistic $\hat{R}$ and stop sampling when $\hat{R}-1<0.05$ for all marginalized parameters and $\hat{R}-1<0.2$ for all 95\% credible intervals.
\end{itemize}

\section{Results}
\label{sec:results}
\vspace{0.2cm}
In this section we present the main results of this study. We first perform a comprehensive validation check of the \textsc{CosmoLike} model for the 3$\times$2pt probes in Section~\ref{sec:code_comparison}, by comparing it with models generated by \textsc{CCL}. Having validated the model, we define in Section~\ref{sec:setup} our numerical experiment setup, and next in Section~\ref{sec:approximation_impact} we use \textsc{CoCoA} to quantify the impact of the Limber approximation, RSD, and choices of nonlinear matter power spectrum on the 3$\times$2pt data vector and the resulting cosmological constraints for both Roman HLIS-Y5 and Rubin LSST-Y10. 

\subsection{Code comparison: \textsc{CosmoLike} vs. \textsc{CCL}} 
\label{sec:code_comparison}
\vspace{0.2cm}
\begin{figure}
  \centering
  \includegraphics[width=\linewidth]{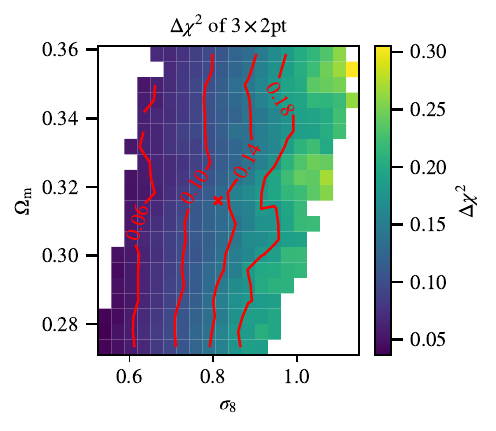}
\vspace{-0.3in}
  \caption{$\Delta\chi^2$ of $3\times2\mathrm{pt}$ simulated by \textsc{CosmoLike} and \textsc{CCL} with varying cosmologies over $\Omega_{\rm m}-\sigma_8$. The red lines with numbers are $\Delta\chi^2$ contours over cosmologies. The red cross represents our fiducial cosmology.}
  \label{fig:chi2-over-cosmology}
\end{figure}

We first compare all the building blocks used to compute the 3$\times$2pt data vectors: the matter power spectrum, the comoving radial distance, the radial kernel, and the lensing efficiency. Next, we combine the components and compare the final 3$\times$2pt data vector. We also investigate whether the two pipelines agree under different assumptions of systematic effects (varying parameters in the IA model) and approximations (RSD and non-Limber) that are used in this analysis. 

For the real-space data vectors with and without fiducial scale cuts, we find,
\begin{align*}
    \Delta\chi_{\mathrm{fid}-\mathrm{cut}}^2(\xi^\pm, \gamma_t, &w, 3\times2\mathrm{pt})_\mathrm{Roman} \\
    &= (0.004, 0.007, 0.052, 0.137),\\
    \Delta\chi^2(\xi^\pm, \gamma_t, &w, 3\times2\mathrm{pt})_\mathrm{Roman} \\
    &= (0.032,0.052,0.064,0.167).
\end{align*}
For the harmonic-space data vectors with and without scale cuts, we find,
\begin{align*}
    \Delta\chi_{\mathrm{fid}-\mathrm{cut}}^2(C^{ss}, C^{gs}, &C^{gg}, 3\times2\mathrm{pt})_\mathrm{Roman} \\
    &= (0.003, 0.005, 0.031, 0.075),\\
    \Delta\chi^2(C^{ss}, C^{gs}, &C^{gg}, 3\times2\mathrm{pt})_\mathrm{Roman} \\
    &= (0.026, 0.062, 0.050, 0.198).
\end{align*}

These values are all small (mostly $\ll$1, especially when scale cuts are applied), demonstrating that the two modeling pipelines agree well with each other in our fiducial settings. In Figure~\ref{fig:Code-comparison-roman-real}, we show one example of the difference in part of the real-space data vector in the Roman baseline setting.

There are a few subtleties in this comparison that we would like to highlight, which are important for future similar comparisons
\begin{itemize}
\item In all these tests, we import the matter power spectrum of \textsc{CosmoLike} into \textsc{CCL} using its \textsc{Pk2D} module, thus the two pipelines use exactly the same matter power spectrum. We do this since both codes wrap around \textsc{CAMB} and the exact implementation of each of the codes is not of interest in this work.

\item When we turn off RSD, the $\Delta\chi^2(3\times2\mathrm{pt})$ increases significantly to $\sim$1 and the scale cuts do not mitigate this effect. This is due to inconsistencies in the implementation of the non-Limber algorithm in galaxy clustering, which we are currently investigating. As $\Delta \chi^2\sim 1$ is still much smaller than the effects of interest that we discuss in Section~\ref{sec:approximation_impact}, we do not attempt to correct for it. But future work may need to reconcile this difference.
\item At the time of this analysis, in \textsc{CosmoLike}, non-Limber is only implemented for galaxy clustering and not for galaxy-galaxy lensing or cosmic shear. We can estimate the effect of this approximation for the two probes using \textsc{CCL}. We find $\Delta\chi^2(\xi^\pm,\gamma_t)_\mathrm{Roman} = (0.21, 0.30)$ for Roman HLIS-Y5 and  $\Delta\chi^2(\xi^\pm,\gamma_t)_\mathrm{Rubin} = (0.56, 1.16)$ for Rubin LSST-Y10. We determine that this level of discrepancy is small enough for us to continue our analysis with the existing \textsc{CosmoLike}.
\item For numerical convergence reasons, the implementation of the non-Limber power spectra requires one to set a maximum multipole $\ell_{\rm max}^{\rm non\text{-}Limber}$ for the non-Limber calculation. This method is used for both \textsc{CosmoLike} and \textsc{CCL}. For multipoles beyond $\ell_{\rm max}^{\rm non\text{-}Limber}$, the calculation defaults back to using the Limber approximation. We set $\ell_{\rm max}^{\rm non\text{-}Limber}=150$ when calculating the non-Limber power spectrum using both \textsc{CosmoLike} and \textsc{CCL} throughout this analysis. The introduction of $\ell_{\rm max}^{\rm non\text{-}Limber}$ results in a slight underestimate of difference in both data vector and constraints when comparing Limber vs. non-Limber later in Section~\ref{sec:approximation_impact}.  
\item Our code-comparison exercise reveals two effects in the real space data vectors that must be included in Stage-IV analyses: the angular bin window function (or, bin-average) and the flat/curved-sky transformation. We describe our tests in Appendix~\ref{sec:realDV}.  Both effects are implemented in \textsc{CoCoA}, so here we mostly point out the impact of the results if the effects are not accounted for.
\end{itemize}

Next, we calculate the data vector $\Delta\chi^2$ between two pipelines with the fiducial scale cut over a range of cosmological parameters to check for potential cosmology-dependent effects. 

Figure~\ref{fig:chi2-over-cosmology} shows the $\Delta\chi^2$ for $3\times2\mathrm{pt}$ as a function of different $\Omega_{\rm m}$ and $\sigma_8$ values. We find that the $\Delta\chi^2$ can rise to around 0.3 when $\sigma_8$ becomes larger though this number is still sufficiently small. We argue that such dependence on cosmology is mainly due to the overall amplitude increase as the $\sigma_8$ rises, illustrating that we should take this cosmology-enhanced $\Delta\chi^2$ into consideration when we compare data vectors at cosmologies that have an overall smaller amplitude of signal.

\subsection{Experiment setup}
\label{sec:setup}
\vspace{0.2cm}
\begin{figure*}
  \centering
  \includegraphics[width=\linewidth]{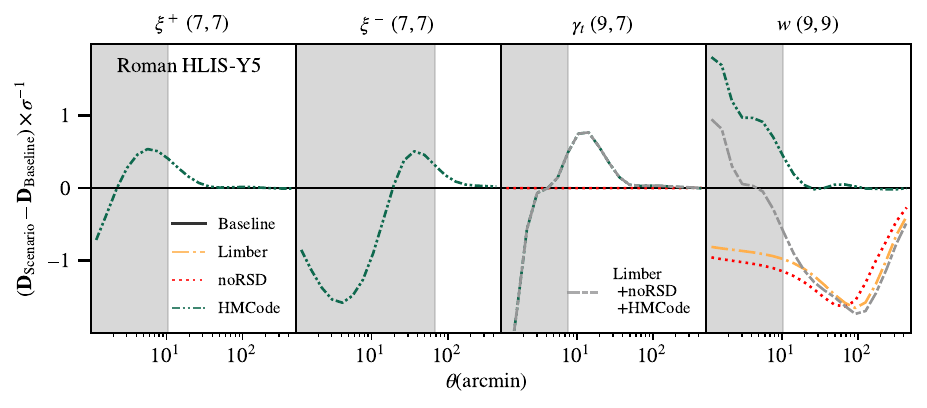}
  \includegraphics[width=\linewidth]{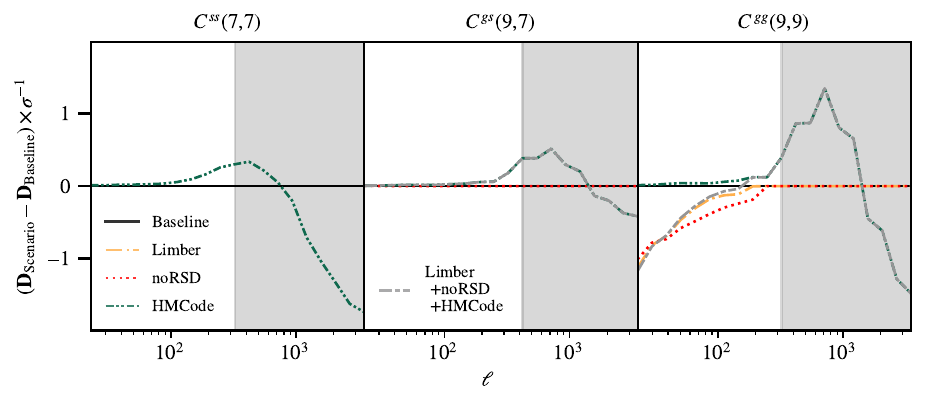}
  \caption{This plot compares data vectors of four scenarios with our \texttt{Baseline}, including \texttt{Limber}, \texttt{noRSD}, \texttt{HMCode}, and \texttt{Limber-noRSD-HMCode}. The difference of data vectors is measured by the diagonal components of the data vector covariance matrix as $(\mathrm{Scenario}-\mathrm{Baseline})/\sqrt{\mathrm{Diag(Cov)}}$, illustrating the significance of bias. The upper figure is for Roman HLIS-Y5 in real space while the lower is for that in harmonic space. The gray shadow represents our DES-like scale cuts. All shown combinations use the highest-redshift bins of the lens- and source-galaxy samples, because the influence of these approximations is more significant in these bins. Note that \texttt{Limber}, \texttt{noRSD} and \texttt{Limber-noRSD-HMCode} are not shown for cosmic shear, since the former two do not affect this probe, while the effect of the last one is identical to that of \texttt{HMCode}. \texttt{Limber} is not implemented for galaxy-galaxy lensing in our analysis and is thus not shown for this probe.}
  \label{fig: main_result_dvs_roman_real_fourier}
\end{figure*}

\begin{figure*}
  \centering
  \includegraphics[width=\linewidth]{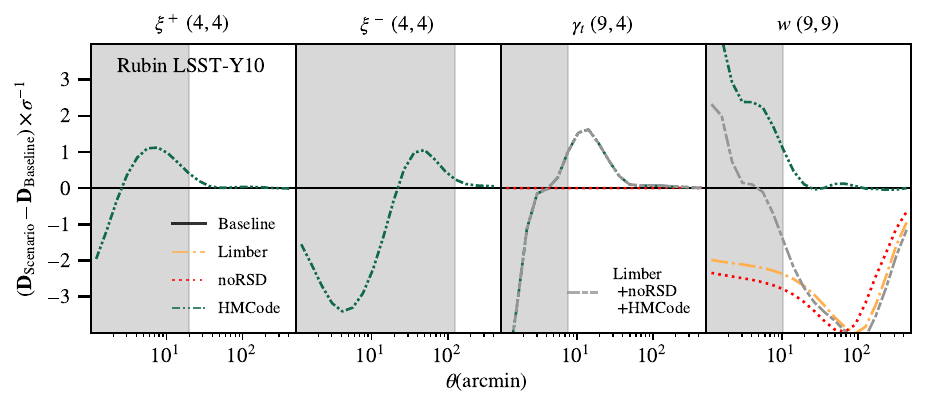}
  \includegraphics[width=\linewidth]{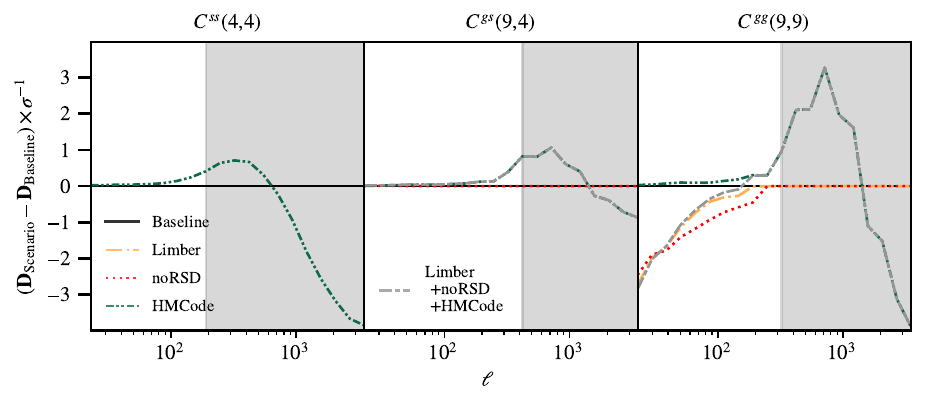}

  \caption{This plot compares data vectors of four scenarios, as described in Figure~\ref{fig: main_result_dvs_roman_real_fourier}, with our \texttt{Baseline}. The data vector difference is measured by the diagonal components of the data vector covariance matrix as $(\mathrm{Scenario}-\mathrm{Baseline})/\sqrt{\mathrm{Diag(Cov)}}$, illustrating the significance of bias. The upper figure is for Rubin LSST-Y10 in the real space while the lower is for that in harmonic space. The gray shadow represents the DES-like scale cuts. All shown combinations use the highest-redshift bins of the lens- and source-galaxy samples, because the influence of these approximations is more significant in these bins.}
  \label{fig: main_result_dvs_lsst_real_fourier}
\end{figure*}

We would like to investigate the three different modeling approximations described in Section~\ref{sec:approx}: the Limber approximation used in projecting 3D power spectra into angular observables, RSD as a correction to the observed galaxy overdensity field, and different models used for the nonlinear matter power spectrum.

For Limber and non-Limber, we investigate their effects only for galaxy clustering in this paper since the Limber approximation does not significantly impact the galaxy-galaxy lensing and cosmic shear data vectors. For RSD, we investigate its influence on galaxy-galaxy lensing and galaxy clustering. We test the impact of the nonlinear matter power spectrum for the full 3$\times$2pt data vector.

To compare these approximations systematically, we define five model setups. 
\begin{itemize}
\item \texttt{Baseline:} this is designed to be the most ``correct'' model setup, where we implement non-Limber for galaxy clustering, RSD for the observed galaxy overdensity field, and \textsc{EuclidEmulator2} for the nonlinear matter power spectrum.
\item \texttt{Limber:} this setup is designed to test the impact of the Limber approximation, so it follows \texttt{Baseline} but uses the Limber approximation for galaxy clustering.
\item \texttt{noRSD:} this setup is designed to test the impact of ignoring RSD, so it follows \texttt{Baseline} but does not include the RSD correction in the observed galaxy overdensity field.
\item \texttt{HMCode:} this setup is designed to test the impact of using a less precise nonlinear matter power spectrum, so it follows \texttt{Baseline} but uses \textsc{HMCode} for the matter power spectrum.
\item \texttt{Limber-noRSD-HMCode:} this setup is designed to test the impact of all three approximations combined, so it uses the Limber approximation and \textsc{HMCode}, and ignores RSD. 
\end{itemize}
By comparing \texttt{Baseline} with the other four setups, we are able to quantify how these approximations affect the data vector simulations and cosmological inference.

\subsection{Impact of approximations}
\label{sec:approximation_impact}
\vspace{0.2cm}
We first simulate data vectors for these setups and quantify the discrepancy that those approximations could introduce to the data vectors. All data vectors are simulated under four scenarios for two surveys, Roman HLIS-Y5 and Rubin LSST-Y10 (Table~\ref{tab:survey_settings}), and for both real and harmonic spaces. These scenarios are noted as Roman-Real, Roman-Harmonic, Rubin-Real, and Rubin-Harmonic. 

\begin{table}
  \centering
  \caption{$\Delta\chi^2$ between data vectors of different scenarios and the baseline, and the significance of bias in the $(\Omega_\mathrm{m}, S_8)$ plane relative to the baseline chain.}
  \label{tab:dv-chi2-bias}
  \vspace{0.3em}
  \begin{tabular}{lcccc}
    \toprule
    & \texttt{Limber} & \texttt{noRSD} & \texttt{HMCode} & \texttt{Limber} \\
    & & & & \texttt{+noRSD+HMCode} \\
    \midrule
    \multicolumn{5}{l}{$\boldsymbol{\Delta\chi^2}$} \\
    DES-Y3         & 9.15   & 8.90   & 0.93  & 12.29  \\
    Roman-Real     & 23.85  & 23.66  & 6.75  & 34.79  \\
    Roman-Harmonic & 13.43  & 16.33  & 1.13  & 15.00  \\
    Rubin-Real     & 131.65 & 129.10 & 34.99 & 190.62 \\
    Rubin-Harmonic & 71.42  & 87.09  & 6.03  & 79.44  \\

    \midrule
    \multicolumn{5}{l}{\textbf{Significance of bias $(\sigma)$}} \\
    Roman-Real     & 0.73 & 0.70 & 0.61 & 0.96 \\
    Roman-Harmonic & 0.75 & 0.75 & 0.56 & 1.12 \\
    Rubin-Real     & 1.79 & 2.05 & 0.98 & 2.76 \\
    Rubin-Harmonic & 1.97 & 2.31 & 1.24 & 3.12 \\
    \bottomrule
  \end{tabular}
\end{table}

Figure~\ref{fig: main_result_dvs_roman_real_fourier} shows the comparison for Roman-Real and Roman-Harmonic for a subset of the data vector, while Figure~\ref{fig: main_result_dvs_lsst_real_fourier} shows the same for Rubin-Real and Rubin-Harmonic. The x-axis demonstrates the scale of the observables which is over either angular scale ($\mathrm{arcmin}$) or harmonic modes ($\ell$), while the y-axis shows the bias significance of the data vectors comparing the scenarios, including  \texttt{Limber}, \texttt{noRSD}, \texttt{HMCode}, and \texttt{Limber-noRSD-HMCode}, with \texttt{Baseline}. The significance of bias is calculated by dividing the difference of data vectors by the square root of the diagonal components of the data vector covariance matrix.

For brevity, we only show one bin combination for each of the data vector types ($\xi_{\pm}$, $\gamma_t$, $w$ and $C^{ss}$, $C^{gs}$, $C^{gg}$) -- we choose to show the highest-redshift bin as an illustration. Because \texttt{Limber} and \texttt{noRSD} demonstrate greater discrepancy relative to \texttt{Baseline} at higher redshift. For \texttt{HMCode}, although it shows a larger difference at lower redshift on small scales, the difference on intermediate scales is somewhat redshift-independent. 

\begin{figure*}
  \centering
  \includegraphics[width=\linewidth]{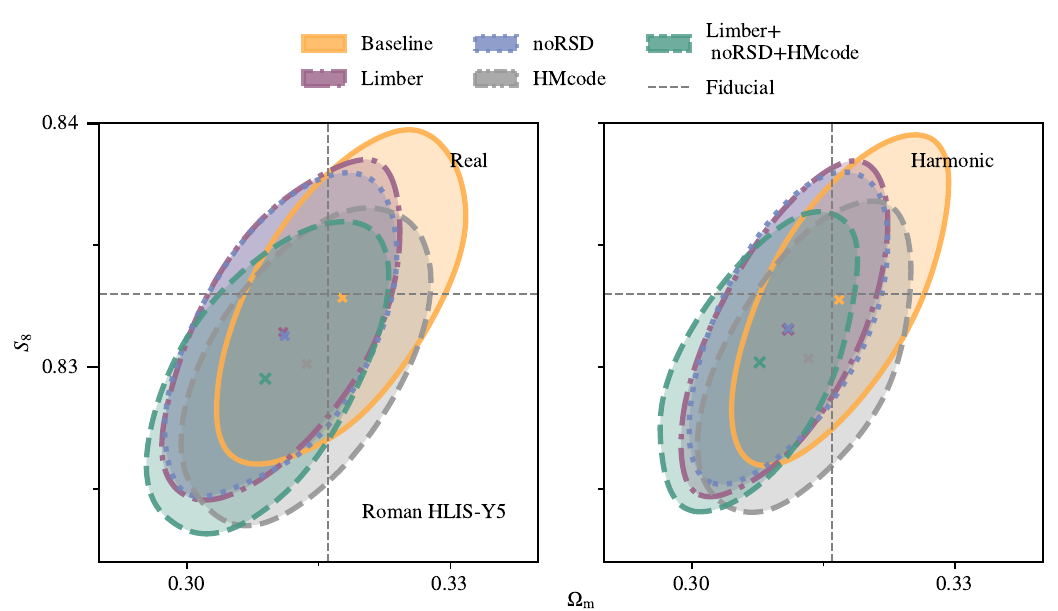}

  \caption{This plot shows $S_8-\Omega_\mathrm{m}$ constraints of five scenarios, including \texttt{Baseline}, \texttt{Limber}, \texttt{noRSD}, \texttt{HMCode}, and \texttt{Limber-noRSD-HMCode}, for survey Roman HLIS-Y5. The contours enclose a 68\% credible region. The colored crosses represent the posterior means of scenarios, while the gray cross-hair represents the parameter value of the fiducial cosmology. }\label{fig:main_result_contours_roman_lcdm}
\end{figure*}

We first look at the cosmic shear panels. Since RSD only influences galaxy-galaxy lensing and galaxy clustering, and non-Limber is only applied to galaxy clustering, the only approximation that can affect cosmic shear is the nonlinear matter power spectrum model. As seen in Figure~\ref{fig: main_result_dvs_roman_real_fourier} and Figure~\ref{fig: main_result_dvs_lsst_real_fourier}, setup \texttt{HMCode} and \texttt{Baseline} show a noticeable difference when $\theta<60'$ for $\xi^+$, $\theta<110'$ for $\xi^-$, and $\ell>100$ for $C_\ell^{ss}$, though most of the affected scales are cut by our fiducial scale cuts. The typical significance of bias to the data vectors introduced by the nonlinear matter power spectrum is within $1\sigma$ for Roman HLIS-Y5 and about $2\sigma$ for Rubin LSST-Y10, while both of them are well controlled under $1\sigma$ after scale cuts.

Next we look at the galaxy-galaxy lensing panels in Figure~\ref{fig: main_result_dvs_roman_real_fourier} and Figure~\ref{fig: main_result_dvs_lsst_real_fourier}. We see that the nonlinear matter power spectrum has the most significant impact on the data vectors, while that of RSD is negligible. This is because RSD is greatly suppressed under the Limber approximation as its $k_\parallel$ mode, defined in Section~\ref{sec: redshift-space-distortion}, is near 0. The influence of the nonlinear matter power spectrum becomes noticeable when $\theta<60'$ and $\ell>100$, and the typical bias significance to data vectors is below $1\sigma$ for Roman HLIS-Y5 and about $1\sigma$ for Rubin LSST-Y10 after the fiducial scale cuts.

\begin{figure*}
  \centering
  \includegraphics[width=\linewidth]{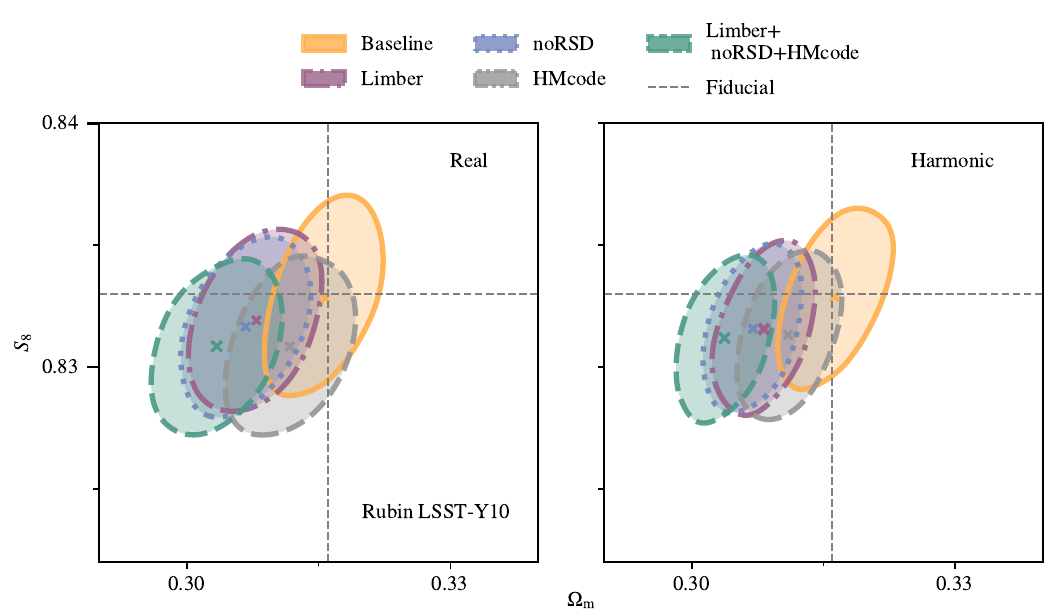}
  \caption{This plot shows $S_8-\Omega_\mathrm{m}$ constraints of five scenarios as described in Figure~\ref{fig:main_result_contours_roman_lcdm} for survey Rubin LSST-Y10. The contours enclose an area corresponding to 68\%  credible region. The colored crosses represent the posterior means of scenarios, while the gray cross-hair represents the input parameter value of fiducial cosmology.}\label{fig:main_result_contours_lsst_lcdm}
\end{figure*}

Finally we examine the panels for galaxy clustering. The impact of the nonlinear matter power spectrum is still significant when $\theta<20'$ and $\ell>100$, which introduces a typical bias below $1\sigma$ for Roman HLIS-Y5 and around $1\sigma$ for Rubin LSST-Y10 after scale cuts. The RSD is important as it introduces a typical $1\sigma$ bias to data vectors for Roman HLIS-Y5 and $2\sigma$ for Rubin LSST-Y10.

The Limber approximation has a similar level of impact as the RSD, but since the Limber approximation strongly suppresses the RSD contribution, this demonstrates that RSD is the dominant effect between the two.

We show the resulting posteriors in the $\Omega_{\rm m}$-$S_8$ plane in Figures~\ref{fig:main_result_contours_roman_lcdm} and \ref{fig:main_result_contours_lsst_lcdm}. These plots allow us to translate the $\Delta\chi^2$ numbers in Table~\ref{tab:dv-chi2-bias} to shifts in cosmological constraints. We summarize the shift of posterior mean between \texttt{Baseline} and other setups in Table~\ref{tab:dv-chi2-bias} and discuss them below.  

First, we show that in all scenarios, the posterior mean of the \texttt{Baseline} setup lies close to the input fiducial values as demonstrated by the orange cross (\texttt{Baseline}) and black cross-hair (Fiducial), respectively, in Figure~\ref{fig:main_result_contours_roman_lcdm} and \ref{fig:main_result_contours_lsst_lcdm}. Numerically, the bias significance of \texttt{Baseline} posterior mean compared to the fiducial value is $0.25\sigma$ in Roman-Real, $0.18\sigma$ in Roman-Harmonic, $0.07\sigma$ in Rubin-Real, and $0.11\sigma$ in Rubin-Harmonic. These numbers mainly quantify the level of the projection effect in our settings. For all cases, we argue that these values are sufficiently small for proceeding the analysis.

Next, we look at the impact of the approximations on biases in cosmological constraints. For \texttt{Limber}, it introduces a bias of $0.73\sigma$ to Roman-Real and a bias of $0.75\sigma$ to Roman-Harmonic, while the number for Rubin LSST-Y10 is larger, at $1.79\sigma$ to Rubin-Real and $1.97\sigma$ to Rubin-Harmonic. We note that the Limber approximation mainly affects data vectors on large scales, where the scale cuts have little impact, as shown in Figures~\ref{fig: main_result_dvs_roman_real_fourier} and ~\ref{fig: main_result_dvs_lsst_real_fourier}. The affected scales of the \texttt{Limber} scenario are retained after the harmonic-space scale cuts, while a small part of them is removed further by the scale cuts in real space. This explains why the bias in harmonic space is larger than that in real space.

For \texttt{noRSD}, it introduces a bias of $0.70\sigma$ to Roman-Real, $0.75\sigma$ to Roman-Harmonic, $2.05\sigma$ to Rubin-Real and $2.31\sigma$ to Rubin-Harmonic. The bias introduced by \texttt{noRSD} is comparable to and even stronger than that of \texttt{Limber}. This is revealed by the red dotted line in panels of galaxy clustering in Figures~\ref{fig: main_result_dvs_roman_real_fourier} and ~\ref{fig: main_result_dvs_lsst_real_fourier} that it has equivalent and even greater amplitude compared to that of the yellow line (\texttt{Limber}). The main impact of \texttt{noRSD} lies on the large scales, same as the \texttt{Limber} case, where the scale cuts have little influence. 

For \texttt{HMCode}, it introduces a bias of $0.61\sigma$ to Roman-Real, $0.56\sigma$ to Roman-Harmonic, $0.98\sigma$ to Rubin-Real, and $1.24\sigma$ to Rubin-Harmonic. These numbers are smaller than those of \texttt{Limber} and \texttt{noRSD}, because the difference in the data vectors due to different nonlinear matter power spectrum prescriptions is mainly on small scales, which can be removed by the scale cuts. 
This can be seen by the green line in Figures~\ref{fig: main_result_dvs_roman_real_fourier} and ~\ref{fig: main_result_dvs_lsst_real_fourier} that scales where it contributes most are covered by our scale cuts.

The impact of all three approximations are found to be larger in the Rubin LSST-Y10 than in the Roman HLIS-Y5 since the former has greater constraining power. Note that we conservatively assume medium tier for Roman in this analysis, though its constraining power can be strongly boosted when assuming wide tier. 

For \texttt{Limber+noRSD+HMCode}, it introduces the greatest bias significance, $0.96\sigma$ to Roman-Real, $1.12\sigma$ to Roman-Harmonic, $2.76\sigma$ to Rubin-Real, and $3.12\sigma$ to Rubin-Harmonic. This demonstrates how much the inferred cosmology could be biased if we neglect all three modeling effects, though the bias introduced by any single modeling component may already be unacceptable. We note that since \texttt{HMCode} mainly biases the data vector on small scales, and \texttt{noRSD} and \texttt{Limber} have similar effects on large scales, their effects do not simply cancel each other. 

We also calculate the Figure of Merit (FoM) in the $\Omega_{\rm m}-S_8$ plane for each survey's \texttt{Baseline} case defined as
$\text{FoM} \equiv 1/(\sqrt{\text{det}(\text{Cov}_{\Omega_{\rm m}-S_8})})$. The FoM is a metric of the constraining power, that a large number of FoM indicates tighter contours and a small number of FoM indicates weaker constraints. 

We find a FoM of $2.9\times10^4$ for Roman-Real, $3.7\times10^4$ for Roman-Harmonic, $9.3\times10^4$ for Rubin-Real, and $1.1\times10^5$ for Rubin-Harmonic. The harmonic space FoM is generally larger than that of real space in our specific scale cut implementation. 

\section{Conclusions}
\label{sec:conclusion}
\vspace{0.2cm}
In this paper, we study the three physical modeling approximations in a standard $3\times2\mathrm{pt}$ cosmology analysis: the Limber approximation, omitting RSD, and imperfect modeling of the nonlinear matter power spectrum. These approximations can bias the data vector, and the resulting bias can propagate into bias in the inferred cosmological parameters. Some of these approximations have been adopted in Stage-III galaxy imaging surveys, where the statistical power is not high enough for the approximations to impact the final analyses. However, when we move on to the Stage-IV surveys such as Roman and Rubin, it is important to revisit these approximations.

We set up numerical experiments forecasting the inference performance for the final 5-year Roman High Latitude Imaging Survey and the final 10-year Rubin Legacy Survey of Space and Time. We use the software package \textsc{CoCoA}, which wraps around the cosmology library \textsc{CosmoLike} to carry out our analysis. We cross-check \textsc{CosmoLike} with the established code base \textsc{CCL} and find excellent agreements between the two codes. 

We next turn on and off each of the three approximations, and observe the difference in the data vectors and the resulting cosmological constraints. We find that using the Limber approximation or not including RSD can result in a $\Delta\chi^2$ of $13\sim24$ for Roman and $71\sim132$ for Rubin, while the cosmological bias induced in the $\Omega_\mathrm{m}-S_8$ plane is about $1\sigma$ for Roman and $2\sigma$ for Rubin. We also find that using a different nonlinear matter power spectrum prescription, i.e. \textsc{HMCode} compared with \textsc{EuclidEmulator2} can result in a $\Delta\chi^2$ of $1\sim7$ for Roman and $6\sim35$ for Rubin, while the bias is about $0.6\sim0.8\sigma$ for Roman and $0.9\sim1.3\sigma$ for Rubin. These numbers can vary under different configurations of the data vector (real or harmonic), scale cuts, and covariance matrix.

Our results have implications for the Stage-IV surveys: 
\begin{enumerate}
\item It is important to optimize the non-Limber algorithm so that it can become more numerically stable and time-efficient. 
\item For RSD, our work studies the linear-order effect and finds a large impact on the data vector and cosmological parameters. This motivates a more realistic RSD model. 
\item The choice of the nonlinear matter power spectrum model is important, and there is an interplay between the importance of the model and the chosen scale cuts, where more aggressive scale cuts increase the sensitivity to such choices. 
\end{enumerate}
We discuss in Appendices~\ref{sec:lens_sample} and ~\ref{app:analysis_choice} how our main results will change when a different lens sample is used and when a different scale cut choice is used.

We imagine several extensions to this work: First, as we have investigated some of the more basic approximations, the next step could be to focus on more advanced modeling components such as the nonlinear galaxy bias, the intrinsic alignment model, and the baryonic feedback model. Second, although we have made an attempt to make our survey configurations more realistic by adopting different lens samples, compared to existing literature, and a DES-like scale cut, there are more aspects in our analysis that could be made more realistic based on what we have learned in the Stage-III surveys, including the redshift distributions of source- and lens-galaxy samples, and the sample selection. Understanding the effect of these simplifications is important. Finally, our work only uses the Roman medium tier for simplicity, while there will also be the data of wide tier, which could significantly increase the statistical power. Investigating how to combine the medium- and wide-tier dataset to maximize the statistical power of the Roman survey is an interesting future direction.    

\section*{Acknowledgments}
\vspace{0.2cm}
We acknowledge support from the ``Maximizing Cosmological Science with the Roman High Latitude Imaging Survey'' Roman Project Infrastructure Team. This work was supported by the NASA ROSES grant 22-ROMAN11-0011, contract number 80NM0024F0012, via a JPL subaward.

We are grateful for the support of the University of Chicago's Research Computing Center for assistance with the calculations carried out in this work.

This work made use of the open-source software packages: \texttt{CoCoA}, \texttt{CosmoLike} \citep{Krause2017, Fang__2020}, \texttt{CosmoCov}, \texttt{Cobaya} \citep{Torrado2021}, \texttt{CAMB} \citep{Lewis_2000}, \texttt{CCL} \citep{Chisari_2019}, \texttt{NumPy} \citep{harris2020array}, \texttt{SciPy} \citep{2020SciPy-NMeth}, \texttt{Matplotlib} \citep{Hunter:2007}.

\bibliographystyle{mnras}
\bibliography{reference}

@ARTICLE{2004PhRvD..70d3009H,
       author = {{Hu}, Wayne and {Jain}, Bhuvnesh},
        title = "{Joint galaxy-lensing observables and the dark energy}",
      journal = {\prd},
         year = 2004,
        month = aug,
       volume = {70},
       number = {4},
          eid = {043009},
        pages = {043009},
          doi = {10.1103/PhysRevD.70.043009},
archivePrefix = {arXiv},
       eprint = {astro-ph/0312395},
 primaryClass = {astro-ph},
       adsurl = {https://ui.adsabs.harvard.edu/abs/2004PhRvD..70d3009H}
}

@misc{sanchezcid2026darkenergysurveyyear,
      title={Dark Energy Survey Year 6 Results: Weak Lensing and Galaxy Clustering Cosmological Analysis Framework}, 
      author={D. Sanchez-Cid and A. Ferté and J. Blazek and S. Samuroff and A. Amon and F. Andrade-Oliveira and J. M. Coloma-Nadal and J. Muir and A. Porredon and J. Prat and N. Weaverdyck and M. Yamamoto and D. Anbajagane and M. R. Becker and P. Carrilho and C. Chang and M. Crocce and G. Giannini and W. d'Assignies and J. DeRose and S. Dodelson and E. Krause and E. Legnani and J. Mena-Fernández and N. MacCrann and A. Pourtsidou and C. Preston and P. Rogozenski and M. Rodriguez-Monroy and R. Rosenfeld and E. Sanchez and I. Sevilla-Noarbe and M. Soares-Santos and C. To and M. A. Troxel and M. Tsedrik and B. Yin and J. Zuntz and T. M. C. Abbott and M. Aguena and S. Allam and O. Alves and S. Avila and D. Bacon and K. Bechtol and E. Bertin and S. Bocquet and D. Brooks and H. Camacho and R. Camilleri and A. Campos and A. Carnero Rosell and J. Carretero and F. J. Castander and R. Cawthon and A. Choi and L. N. da Costa and M. E. da Silva Pereira and T. M. Davis and J. De Vicente and S. Desai and C. Doux and A. Drlica-Wagner and T. Eifler and J. Elvin-Poole and S. Everett and A. E. Evrard and B. Flaugher and P. Fosalba and J. Frieman and J. García-Bellido and M. Gatti and E. Gaztanaga and P. Giles and K. Glazebrook and D. Gruen and G. Gutierrez and I. Harrison and K. Herner and S. R. Hinton and D. L. Hollowood and K. Honscheid and D. Huterer and B. Jain and D. J. James and N. Jeffrey and T. Kacprzak and K. Kuehn and O. Lahav and S. Lee and J. L. Marshall and F. Menanteau and R. Miquel and J. J. Mohr and J. Myles and R. C. Nichol and R. L. C. Ogando and A. Palmese and M. Paterno and W. J. Percival and A. A. Plazas Malagón and M. Raveri and A. Roodman and C. Sánchez and T. Schutt and E. Sheldon and N. Sherman and T. Shin and M. Smith and E. Suchyta and M. E. C. Swanson and M. Tabbutt and G. Tarle and D. Thomas and D. L. Tucker and V. Vikram and A. R. Walker and B. Yanny},
      year={2026},
      eprint={2601.14859},
      archivePrefix={arXiv},
      primaryClass={astro-ph.CO},
      url={https://arxiv.org/abs/2601.14859}, 
}

@article{Secco_2022,
   title={Dark Energy Survey Year 3 results: Cosmology from cosmic shear and robustness to modeling uncertainty},
   volume={105},
   ISSN={2470-0029},
   url={http://dx.doi.org/10.1103/PhysRevD.105.023515},
   DOI={10.1103/physrevd.105.023515},
   number={2},
   journal={Physical Review D},
   publisher={American Physical Society (APS)},
   author={Secco, L. F. and Samuroff, S. and Krause, E. and Jain, B. and Blazek, J. and Raveri, M. and Campos, A. and Amon, A. and Chen, A. and Doux, C. and Choi, A. and Gruen, D. and Bernstein, G. M. and Chang, C. and DeRose, J. and Myles, J. and Ferté, A. and Lemos, P. and Huterer, D. and Prat, J. and Troxel, M. A. and MacCrann, N. and Liddle, A. R. and Kacprzak, T. and Fang, X. and Sánchez, C. and Pandey, S. and Dodelson, S. and Chintalapati, P. and Hoffmann, K. and Alarcon, A. and Alves, O. and Andrade-Oliveira, F. and Baxter, E. J. and Bechtol, K. and Becker, M. R. and Brandao-Souza, A. and Camacho, H. and Carnero Rosell, A. and Carrasco Kind, M. and Cawthon, R. and Cordero, J. P. and Crocce, M. and Davis, C. and Di Valentino, E. and Drlica-Wagner, A. and Eckert, K. and Eifler, T. F. and Elidaiana, M. and Elsner, F. and Elvin-Poole, J. and Everett, S. and Fosalba, P. and Friedrich, O. and Gatti, M. and Giannini, G. and Gruendl, R. A. and Harrison, I. and Hartley, W. G. and Herner, K. and Huang, H. and Huff, E. M. and Jarvis, M. and Jeffrey, N. and Kuropatkin, N. and Leget, P.-F. and Muir, J. and Mccullough, J. and Navarro Alsina, A. and Omori, Y. and Park, Y. and Porredon, A. and Rollins, R. and Roodman, A. and Rosenfeld, R. and Ross, A. J. and Rykoff, E. S. and Sanchez, J. and Sevilla-Noarbe, I. and Sheldon, E. S. and Shin, T. and Troja, A. and Tutusaus, I. and Varga, T. N. and Weaverdyck, N. and Wechsler, R. H. and Yanny, B. and Yin, B. and Zhang, Y. and Zuntz, J. and Abbott, T. M. C. and Aguena, M. and Allam, S. and Annis, J. and Bacon, D. and Bertin, E. and Bhargava, S. and Bridle, S. L. and Brooks, D. and Buckley-Geer, E. and Burke, D. L. and Carretero, J. and Costanzi, M. and da Costa, L. N. and De Vicente, J. and Diehl, H. T. and Dietrich, J. P. and Doel, P. and Ferrero, I. and Flaugher, B. and Frieman, J. and García-Bellido, J. and Gaztanaga, E. and Gerdes, D. W. and Giannantonio, T. and Gschwend, J. and Gutierrez, G. and Hinton, S. R. and Hollowood, D. L. and Honscheid, K. and Hoyle, B. and James, D. J. and Jeltema, T. and Kuehn, K. and Lahav, O. and Lima, M. and Lin, H. and Maia, M. A. G. and Marshall, J. L. and Martini, P. and Melchior, P. and Menanteau, F. and Miquel, R. and Mohr, J. J. and Morgan, R. and Ogando, R. L. C. and Palmese, A. and Paz-Chinchón, F. and Petravick, D. and Pieres, A. and Plazas Malagón, A. A. and Rodriguez-Monroy, M. and Romer, A. K. and Sanchez, E. and Scarpine, V. and Schubnell, M. and Scolnic, D. and Serrano, S. and Smith, M. and Soares-Santos, M. and Suchyta, E. and Swanson, M. E. C. and Tarle, G. and Thomas, D. and To, C. and },
   year={2022},
   month=Jan }

@ARTICLE{Arico2021,
       author = {{Aric{\`o}}, Giovanni and {Angulo}, Raul E. and {Contreras}, Sergio and {Ondaro-Mallea}, Lurdes and {Pellejero-Iba{\~n}ez}, Marcos and {Zennaro}, Matteo},
        title = "{The BACCO simulation project: a baryonification emulator with neural networks}",
      journal = {\mnras},
         year = 2021,
        month = sep,
       volume = {506},
       number = {3},
        pages = {4070-4082},
          doi = {10.1093/mnras/stab1911},
archivePrefix = {arXiv},
       eprint = {2011.15018},
 primaryClass = {astro-ph.CO},
       adsurl = {https://ui.adsabs.harvard.edu/abs/2021MNRAS.506.4070A}
}

@ARTICLE{Schneider2015,
       author = {{Schneider}, Aurel and {Teyssier}, Romain},
        title = "{A new method to quantify the effects of baryons on the matter power spectrum}",
      journal = {\jcap},
         year = 2015,
        month = dec,
       volume = {2015},
       number = {12},
        pages = {049-049},
          doi = {10.1088/1475-7516/2015/12/049},
archivePrefix = {arXiv},
       eprint = {1510.06034},
 primaryClass = {astro-ph.CO},
       adsurl = {https://ui.adsabs.harvard.edu/abs/2015JCAP...12..049S}
}

@ARTICLE{RS2025,
       author = {{R.~S.}, Pranjal and {Krause}, Elisabeth and {Dolag}, Klaus and {Benabed}, Karim and {Eifler}, Tim and {Ay{\c{c}}oberry}, Emma and {Dubois}, Yohan},
        title = "{Impact of cosmology dependence of baryonic feedback in weak lensing}",
      journal = {\jcap},
         year = 2025,
        month = mar,
       volume = {2025},
       number = {3},
          eid = {041},
        pages = {041},
          doi = {10.1088/1475-7516/2025/03/041},
archivePrefix = {arXiv},
       eprint = {2410.21980},
 primaryClass = {astro-ph.CO},
       adsurl = {https://ui.adsabs.harvard.edu/abs/2025JCAP...03..041R}
}

@article{Chisari2018,
    author = {Chisari, Nora Elisa and Richardson, Mark L. A. and
              Devriendt, Julien and Dubois, Yohan and Schneider, Aurel and
              Le Brun, Amandine M. C. and Beckmann, Ricarda S. and
              Peirani, Sebastien and Slyz, Adrianne and Pichon, Christophe},
    title = {The impact of baryons on the matter power spectrum from the
             {Horizon-AGN} cosmological hydrodynamical simulation},
    journal = {Monthly Notices of the Royal Astronomical Society},
    year = {2018},
    volume = {480},
    pages = {3962--3977},
    doi = {10.1093/mnras/sty2093},
    archivePrefix = {arXiv},
    eprint = {1801.08559},
    primaryClass = {astro-ph.CO}
}

@ARTICLE{Cao2026,
       author = {{Cao}, Kaili and {Weinberg}, David H. and {Miranda}, Vivian and {Dalal}, Nihar and {Eifler}, Tim and {Xu}, Jiachuan and {Bowden}, Haley},
        title = "{Fisher Forecasts for Cosmological Yields from $3\!\times\!2$pt Analysis of the Roman Space Telescope High Latitude Imaging Survey}",
      journal = {arXiv e-prints},
         year = 2026,
        month = jan,
          eid = {arXiv:2601.00438},
        pages = {arXiv:2601.00438},
          doi = {10.48550/arXiv.2601.00438},
archivePrefix = {arXiv},
       eprint = {2601.00438},
 primaryClass = {astro-ph.CO},
       adsurl = {https://ui.adsabs.harvard.edu/abs/2026arXiv260100438C}
}

@article{Takahashi2012,
    author = {Takahashi, Ryuichi and Sato, Masanori and Nishimichi, Takahiro
              and Taruya, Atsushi and Oguri, Masamune},
    title = {Revising the {Halofit} Model for the Nonlinear Matter
             Power Spectrum},
    journal = {The Astrophysical Journal},
    year = {2012},
    volume = {761},
    pages = {152},
    doi = {10.1088/0004-637X/761/2/152},
    archivePrefix = {arXiv},
    eprint = {1208.2701},
    primaryClass = {astro-ph.CO}
}

@article{EuclidEmulator2_2021,
    author = {{Euclid Collaboration} and Knabenhans, M. and Stadel, J. and
              Potter, D. and Dakin, J. and Hannestad, S. and Tram, T. and
              Marelli, S. and Schneider, A. and Teyssier, R. and
              Fosalba, P. and others},
    title = {{Euclid} preparation: {IX}. {EuclidEmulator2} -- power spectrum
             emulation with massive neutrinos and self-consistent dark energy
             perturbations},
    journal = {Monthly Notices of the Royal Astronomical Society},
    year = {2021},
    volume = {505},
    pages = {2840--2869},
    doi = {10.1093/mnras/stab1366},
    archivePrefix = {arXiv},
    eprint = {2010.11288},
    primaryClass = {astro-ph.CO}
}

@article{HirataSeljak2004,
    author = {Hirata, Christopher M. and Seljak, Uro{\v{s}}},
    title = {Intrinsic alignment-lensing interference as a contaminant
             of cosmic shear},
    journal = {Physical Review D},
    year = {2004},
    volume = {70},
    pages = {063526},
    doi = {10.1103/PhysRevD.70.063526},
    archivePrefix = {arXiv},
    eprint = {astro-ph/0406096},
    primaryClass = {astro-ph}
}

@article{BridleKing2007,
    author = {Bridle, Sarah and King, Lindsay},
    title = {Dark energy constraints from cosmic shear power spectra:
             impact of intrinsic alignments on photometric redshift
             requirements},
    journal = {New Journal of Physics},
    year = {2007},
    volume = {9},
    pages = {444},
    doi = {10.1088/1367-2630/9/12/444},
    archivePrefix = {arXiv},
    eprint = {0705.0166},
    primaryClass = {astro-ph}
}

@article{Tan_2023,
   title={Assessing theoretical uncertainties for cosmological constraints from weak lensing surveys},
   volume={522},
   ISSN={1365-2966},
   url={http://dx.doi.org/10.1093/mnras/stad1142},
   DOI={10.1093/mnras/stad1142},
   number={3},
   journal={Monthly Notices of the Royal Astronomical Society},
   publisher={Oxford University Press (OUP)},
   author={Tan, Ting and Zürcher, Dominik and Fluri, Janis and Refregier, Alexandre and Tarsitano, Federica and Kacprzak, Tomasz},
   year={2023},
   month=Apr, pages={3766–3783} }

@ARTICLE{Zhang2025_HSC3x2pt,
       author = {{Zhang}, Tianqing and {Li}, Xiangchong and {Sugiyama}, Sunao and {Mandelbaum}, Rachel and {More}, Surhud and {Dalal}, Roohi and {Kannawadi}, Arun and {Miyatake}, Hironao and {Nishizawa}, Atsushi J. and {Nishimichi}, Takahiro and {Oguri}, Masamune and {Osato}, Ken and {Rau}, Markus M. and {Shirasaki}, Masato and {Sunayama}, Tomomi and {Takada}, Masahiro},
        title = "{Cosmology and source redshift constraints from galaxy clustering and tomographic weak lensing with HSC Y3 and SDSS using the point-mass correction model}",
      journal = {\prd},
         year = 2026,
        month = may,
       volume = {113},
       number = {10},
          eid = {103530},
        pages = {103530},
          doi = {10.1103/hskc-8792},
archivePrefix = {arXiv},
       eprint = {2507.01386},
 primaryClass = {astro-ph.CO},
       adsurl = {https://ui.adsabs.harvard.edu/abs/2026PhRvD.113j3530Z}
}

@article{Heymans2021_KiDS1000_3x2pt,
    author = {Heymans, Catherine and Tr{\"o}ster, Tilman and Asgari, Marika and
              Blake, Chris and Hildebrandt, Hendrik and Joachimi, Benjamin and
              Kuijken, Konrad and Lin, Chieh-An and S{\'a}nchez, Ariel G. and
              van den Busch, Jan Luca and Wright, Angus H. and Amon, Alexandra and
              Bilicki, Maciej and de Jong, Jelte and Crocce, Martin and
              Dvornik, Andrej and Erben, Thomas and Fortuna, Maria Cristina and
              Getman, Fedor and Giblin, Benjamin and Glazebrook, Karl and
              Hoekstra, Henk and Joudaki, Shahab and Kannawadi, Arun and
              K{\"o}hlinger, Fabian and Lidman, Chris and Miller, Lance and
              Napolitano, Nicola R. and Parkinson, David and Schneider, Peter and
              Shan, HuanYuan and Valentijn, Edwin and Verdoes Kleijn, Gijs and
              Wolf, Christian},
    title = {{KiDS-1000} Cosmology: Multi-probe weak gravitational lensing and
             spectroscopic galaxy clustering constraints},
    journal = {Astronomy \& Astrophysics},
    volume = {646},
    pages = {A140},
    year = {2021},
    doi = {10.1051/0004-6361/202039063},
    archivePrefix = {arXiv},
    eprint = {2007.15632},
    primaryClass = {astro-ph.CO}
}

@ARTICLE{DESY63x2pt2026,
       author = {{DES Collaboration} and {Abbott}, T.~M.~C. and {Adamow}, M. and {Aguena}, M. and {Alarcon}, A. and {Allam}, S.~S. and {Alves}, O. and {Amon}, A. and {Anbajagane}, D. and {Andrade-Oliveira}, F. and {Avila}, S. and {Bacon}, D. and {Baxter}, E.~J. and {Beas-Gonzalez}, J. and {Bechtol}, K. and {Becker}, M.~R. and {Bernstein}, G.~M. and {Bertin}, E. and {Blazek}, J. and {Bocquet}, S. and {Brooks}, D. and {Brout}, D. and {Camacho}, H. and {Camacho-Ciurana}, G. and {Camilleri}, R. and {Campailla}, G. and {Campos}, A. and {Carnero Rosell}, A. and {Carrasco Kind}, M. and {Carretero}, J. and {Carrilho}, P. and {Castander}, F.~J. and {Cawthon}, R. and {Chang}, C. and {Choi}, A. and {Coloma-Nadal}, J.~M. and {Costanzi}, M. and {Crocce}, M. and {d'Assignies}, W. and {da Costa}, L.~N. and {da Silva Pereira}, M.~E. and {Davis}, T.~M. and {De Vicente}, J. and {DeRose}, J. and {Diehl}, H.~T. and {Dodelson}, S. and {Doel}, P. and {Doux}, C. and {Drlica-Wagner}, A. and {Eifler}, T.~F. and {Elvin-Poole}, J. and {Estrada}, J. and {Everett}, S. and {Evrard}, A.~E. and {Fang}, J. and {Farahi}, A. and {Fert{\'e}}, A. and {Flaugher}, B. and {Fosalba}, P. and {Frieman}, J. and {Garc{\'\i}a-Bellido}, J. and {Gatti}, M. and {Gaztanaga}, E. and {Giannini}, G. and {Giles}, P. and {Glazebrook}, K. and {Gorsuch}, M. and {Gruen}, D. and {Gruendl}, R.~A. and {Gschwend}, J. and {Gutierrez}, G. and {Harrison}, I. and {Hartley}, W.~G. and {Henning}, E. and {Herner}, K. and {Hinton}, S.~R. and {Hollowood}, D.~L. and {Honscheid}, K. and {Huff}, E.~M. and {Huterer}, D. and {Jain}, B. and {James}, D.~J. and {Jarvis}, M. and {Jeffrey}, N. and {Jeltema}, T. and {Kacprzak}, T. and {Kent}, S. and {Kovacs}, A. and {Krause}, E. and {Kron}, R. and {Kuehn}, K. and {Lahav}, O. and {Lee}, S. and {Legnani}, E. and {Lidman}, C. and {Lin}, H. and {MacCrann}, N. and {Manera}, M. and {Manning}, T. and {Marshall}, J.~L. and {Mau}, S. and {McCullough}, J. and {Mena-Fern{\'a}ndez}, J. and {Menanteau}, F. and {Miquel}, R. and {Mohr}, J.~J. and {Muir}, J. and {Myles}, J. and {Nichol}, R.~C. and {Nord}, B. and {O'Donnell}, J.~H. and {Ogando}, R.~L.~C. and {Palmese}, A. and {Paterno}, M. and {Peoples}, J. and {Percival}, W.~J. and {Petravick}, D. and {Pieres}, A. and {Plazas Malag{\'o}n}, A.~A. and {Porredon}, A. and {Pourtsidou}, A. and {Prat}, J. and {Preston}, C. and {Raveri}, M. and {Riquelme}, W. and {Rodriguez-Monroy}, M. and {Rogozenski}, P. and {Romer}, A.~K. and {Roodman}, A. and {Rosenfeld}, R. and {Ross}, A.~J. and {Rozo}, E. and {Rykoff}, E.~S. and {Samuroff}, S. and {S{\'a}nchez}, C. and {Sanchez}, E. and {Sanchez Cid}, D. and {Schutt}, T. and {Sevilla-Noarbe}, I. and {Sheldon}, E. and {Sherman}, N. and {Shin}, T. and {Smith}, M. and {Soares-Santos}, M. and {Suchyta}, E. and {Swanson}, M.~E.~C. and {Tabbutt}, M. and {Tarle}, G. and {Thomas}, D. and {To}, C. and {Tong}, A. and {Toribio San Cipriano}, L. and {Troxel}, M.~A. and {Tsedrik}, M. and {Tucker}, D.~L. and {Vikram}, V. and {Walker}, A.~R. and {Weaverdyck}, N. and {Wechsler}, R.~H. and {Weinberg}, D.~H. and {Weller}, J. and {Wetzell}, V. and {Whyley}, A. and {Wilkinson}, R.~D. and {Wiseman}, P. and {Wu}, H.-Y. and {Yamamoto}, M. and {Yanny}, B. and {Yin}, B. and {Zacharegkas}, G. and {Zhang}, Y. and {Zuntz}, J.},
        title = "{Dark Energy Survey Year 6 Results: Cosmological Constraints from Galaxy Clustering and Weak Lensing}",
      journal = {arXiv e-prints},
         year = 2026,
        month = jan,
          eid = {arXiv:2601.14559},
        pages = {arXiv:2601.14559},
          doi = {10.48550/arXiv.2601.14559},
archivePrefix = {arXiv},
       eprint = {2601.14559},
 primaryClass = {astro-ph.CO},
       adsurl = {https://ui.adsabs.harvard.edu/abs/2026arXiv260114559D}
}

@misc{krause2017darkenergysurveyyear,
      title={Dark Energy Survey Year 1 Results: Multi-Probe Methodology and Simulated Likelihood Analyses}, 
      author={E. Krause and T. F. Eifler and J. Zuntz and O. Friedrich and M. A. Troxel and S. Dodelson and J. Blazek and L. F. Secco and N. MacCrann and E. Baxter and C. Chang and N. Chen and M. Crocce and J. DeRose and A. Ferte and N. Kokron and F. Lacasa and V. Miranda and Y. Omori and A. Porredon and R. Rosenfeld and S. Samuroff and M. Wang and R. H. Wechsler and T. M. C. Abbott and F. B. Abdalla and S. Allam and J. Annis and K. Bechtol and A. Benoit-Levy and G. M. Bernstein and D. Brooks and D. L. Burke and D. Capozzi and M. Carrasco Kind and J. Carretero and C. B. D'Andrea and L. N. da Costa and C. Davis and D. L. DePoy and S. Desai and H. T. Diehl and J. P. Dietrich and A. E. Evrard and B. Flaugher and P. Fosalba and J. Frieman and J. Garcia-Bellido and E. Gaztanaga and T. Giannantonio and D. Gruen and R. A. Gruendl and J. Gschwend and G. Gutierrez and K. Honscheid and D. J. James and T. Jeltema and K. Kuehn and S. Kuhlmann and O. Lahav and M. Lima and M. A. G. Maia and M. March and J. L. Marshall and P. Martini and F. Menanteau and R. Miquel and R. C. Nichol and A. A. Plazas and A. K. Romer and E. S. Rykoff and E. Sanchez and V. Scarpine and R. Schindler and M. Schubnell and I. Sevilla-Noarbe and M. Smith and M. Soares-Santos and F. Sobreira and E. Suchyta and M. E. C. Swanson and G. Tarle and D. L. Tucker and V. Vikram and A. R. Walker and J. Weller},
      year={2017},
      eprint={1706.09359},
      archivePrefix={arXiv},
      primaryClass={astro-ph.CO},
      url={https://arxiv.org/abs/1706.09359}, 
}

@article{porredon2021,
  title = {Dark Energy Survey Year 3 results: Optimizing the lens sample in a combined galaxy clustering and galaxy-galaxy lensing analysis},
  author = {Porredon, A. and Crocce, M. and Fosalba, P. and Elvin-Poole, J. and Carnero Rosell, A. and Cawthon, R. and Eifler, T. F. and Fang, X. and Ferrero, I. and Krause, E. and MacCrann, N. and Weaverdyck, N. and Abbott, T. M. C. and Aguena, M. and Allam, S. and Amon, A. and Avila, S. and Bacon, D. and Bertin, E. and Bhargava, S. and Bridle, S. L. and Brooks, D. and Carrasco Kind, M. and Carretero, J. and Castander, F. J. and Choi, A. and Costanzi, M. and da Costa, L. N. and Pereira, M. E. S. and De Vicente, J. and Desai, S. and Diehl, H. T. and Doel, P. and Drlica-Wagner, A. and Eckert, K. and Fert\'e, A. and Flaugher, B. and Frieman, J. and Garc\'{\i}a-Bellido, J. and Gaztanaga, E. and Gerdes, D. W. and Giannantonio, T. and Gruen, D. and Gruendl, R. A. and Gschwend, J. and Gutierrez, G. and Hartley, W. G. and Hinton, S. R. and Hollowood, D. L. and Honscheid, K. and Hoyle, B. and James, D. J. and Jarvis, M. and Kuehn, K. and Kuropatkin, N. and Maia, M. A. G. and Marshall, J. L. and Menanteau, F. and Miquel, R. and Morgan, R. and Palmese, A. and Pandey, S. and Paz-Chinch\'on, F. and Plazas, A. A. and Rodriguez-Monroy, M. and Roodman, A. and Samuroff, S. and Sanchez, E. and Scarpine, V. and Serrano, S. and Sevilla-Noarbe, I. and Smith, M. and Soares-Santos, M. and Suchyta, E. and Swanson, M. E. C. and Tarle, G. and To, C. and Varga, T. N. and Weller, J. and Wilkinson, R. D.},
  collaboration = {DES Collaboration},
  journal = {Phys. Rev. D},
  volume = {103},
  issue = {4},
  pages = {043503},
  numpages = {26},
  year = {2021},
  month = {Feb},
  publisher = {American Physical Society},
  doi = {10.1103/PhysRevD.103.043503},
  url = {https://link.aps.org/doi/10.1103/PhysRevD.103.043503}
}

@techreport{ROTAC_2025_FinalReport,
  title        = {Roman Observations Time Allocation Committee: Final Report and Recommendations},
  author       = {Zasowski, Gail and Jha, Saurabh W. and Chomiuk, Laura and Fan, Xiaohui and Hickox, Ryan and Huber, Dan and Kerins, Eamonn and Kobulnicky, Chip and Lauer, Tod and Sako, Masao and Shapley, Alice and Stephens, Denise and Weinberg, David and Williams, Ben},
  institution  = {Roman Observations Time Allocation Committee},
  year         = {2025},
  month        = apr,
  date         = {2025-04-24},
  url          = {https://roman.gsfc.nasa.gov/science/ccs/ROTAC-Report-20250424-v1.pdf},
  note         = {Ex officio: Lee Armus, Thomas Barclay, Ori Fox, Jeff Kruk, Patrick Lowrance, Julie McEnery, and Kristen McQuinn}
}

@article{Eifler_2021,
   title={Cosmology with the Roman Space Telescope – multiprobe strategies},
   volume={507},
   ISSN={1365-2966},
   url={http://dx.doi.org/10.1093/mnras/stab1762},
   DOI={10.1093/mnras/stab1762},
   number={2},
   journal={Monthly Notices of the Royal Astronomical Society},
   publisher={Oxford University Press (OUP)},
   author={Eifler, Tim and Miyatake, Hironao and Krause, Elisabeth and Heinrich, Chen and Miranda, Vivian and Hirata, Christopher and Xu, Jiachuan and Hemmati, Shoubaneh and Simet, Melanie and Capak, Peter and Choi, Ami and Doré, Olivier and Doux, Cyrille and Fang, Xiao and Hounsell, Rebekah and Huff, Eric and Huang, Hung-Jin and Jarvis, Mike and Kruk, Jeffrey and Masters, Dan and Rozo, Eduardo and Scolnic, Dan and Spergel, David N and Troxel, Michael and von der Linden, Anja and Wang, Yun and Weinberg, David H and Wenzl, Lukas and Wu, Hao-Yi},
   year={2021},
   month=jul, pages={1746–1761} }

@misc{boruah2024machinelearninglsst3x2pt,
      title={Machine Learning LSST 3x2pt analyses -- forecasting the impact of systematics on cosmological constraints using neural networks}, 
      author={Supranta S. Boruah and Tim Eifler and Vivian Miranda and Elyas Farah and Jay Motka and Elisabeth Krause and Xiao Fang and Paul Rogozenski and The LSST Dark Energy Science Collaboration},
      year={2024},
      eprint={2403.11797},
      archivePrefix={arXiv},
      primaryClass={astro-ph.CO},
      url={https://arxiv.org/abs/2403.11797}, 
}

@ARTICLE{Krause2017,
       author = {{Krause}, Elisabeth and {Eifler}, Tim},
        title = "{cosmolike - cosmological likelihood analyses for photometric galaxy surveys}",
      journal = {\mnras},
         year = 2017,
        month = sep,
       volume = {470},
       number = {2},
        pages = {2100-2112},
          doi = {10.1093/mnras/stx1261},
archivePrefix = {arXiv},
       eprint = {1601.05779},
 primaryClass = {astro-ph.CO},
       adsurl = {https://ui.adsabs.harvard.edu/abs/2017MNRAS.470.2100K}
}

@ARTICLE{Torrado2021,
       author = {{Torrado}, Jes{\'u}s and {Lewis}, Antony},
        title = "{Cobaya: code for Bayesian analysis of hierarchical physical models}",
      journal = {\jcap},
         year = 2021,
        month = may,
       volume = {2021},
       number = {5},
          eid = {057},
        pages = {057},
          doi = {10.1088/1475-7516/2021/05/057},
archivePrefix = {arXiv},
       eprint = {2005.05290},
 primaryClass = {astro-ph.IM},
       adsurl = {https://ui.adsabs.harvard.edu/abs/2021JCAP...05..057T}
}

@article{Leonard_2023,
   title={The N5K Challenge: Non-Limber Integration for LSST Cosmology},
   volume={6},
   ISSN={2565-6120},
   url={http://dx.doi.org/10.21105/astro.2212.04291},
   DOI={10.21105/astro.2212.04291},
   journal={The Open Journal of Astrophysics},
   publisher={Maynooth University},
   author={Leonard, C. Danielle and Ferreira, Tassia and Fang, Xiao and Reischke, Robert and Schoeneberg, Nils and Tröster, Tilman and Alonso, David and Campagne, Jean-Eric and Lanusse, François and Slosar, Anže and Ishak, Mustapha},
   year={2023},
   month=feb }

@misc{thelsstdarkenergysciencecollaboration2021lsstdarkenergyscience,
      title={The LSST Dark Energy Science Collaboration (DESC) Science Requirements Document}, 
      author={The LSST Dark Energy Science Collaboration and Rachel Mandelbaum and Tim Eifler and Renée Hložek and Thomas Collett and Eric Gawiser and Daniel Scolnic and David Alonso and Humna Awan and Rahul Biswas and Jonathan Blazek and Patricia Burchat and Nora Elisa Chisari and Ian Dell'Antonio and Seth Digel and Josh Frieman and Daniel A. Goldstein and Isobel Hook and Željko Ivezić and Steven M. Kahn and Sowmya Kamath and David Kirkby and Thomas Kitching and Elisabeth Krause and Pierre-François Leget and Philip J. Marshall and Joshua Meyers and Hironao Miyatake and Jeffrey A. Newman and Robert Nichol and Eli Rykoff and F. Javier Sanchez and Anže Slosar and Mark Sullivan and M. A. Troxel},
      year={2021},
      eprint={1809.01669},
      archivePrefix={arXiv},
      primaryClass={astro-ph.CO},
      url={https://arxiv.org/abs/1809.01669}, 
}

@article{Fang_2020,
   title={Beyond Limber: efficient computation of angular power spectra for galaxy clustering and weak lensing},
   volume={2020},
   ISSN={1475-7516},
   url={http://dx.doi.org/10.1088/1475-7516/2020/05/010},
   DOI={10.1088/1475-7516/2020/05/010},
   number={05},
   journal={Journal of Cosmology and Astroparticle Physics},
   publisher={IOP Publishing},
   author={Fang, Xiao and Krause, Elisabeth and Eifler, Tim and MacCrann, Niall},
   year={2020},
   month=may, pages={010–010} }

@article{Chisari_2019,
doi = {10.3847/1538-4365/ab1658},
url = {https://doi.org/10.3847/1538-4365/ab1658},
year = {2019},
month = {may},
publisher = {The American Astronomical Society},
volume = {242},
number = {1},
pages = {2},
author = {Chisari, Nora Elisa and Alonso, David and Krause, Elisabeth and Leonard, C. Danielle and Bull, Philip and Neveu, Jérémy and Villarreal, Antonio and Singh, Sukhdeep and McClintock, Thomas and Ellison, John and Du, Zilong and Zuntz, Joe and Mead, Alexander and Joudaki, Shahab and Lorenz, Christiane S. and Tröster, Tilman and Sanchez, Javier and Lanusse, Francois and Ishak, Mustapha and Hlozek, Renée and Blazek, Jonathan and Campagne, Jean-Eric and Almoubayyed, Husni and Eifler, Tim and Kirby, Matthew and Kirkby, David and Plaszczynski, Stéphane and Slosar, Anže and Vrastil, Michal and Wagoner, Erika L. and (LSST Dark Energy Science Collaboration)},
title = {Core Cosmology Library: Precision Cosmological Predictions for LSST},
journal = {The Astrophysical Journal Supplement Series}
}

@article{Fang_2021,
   title={Cosmology from clustering, cosmic shear, CMB lensing, and cross correlations: combining Rubin observatory and Simons Observatory},
   volume={509},
   ISSN={1365-2966},
   url={http://dx.doi.org/10.1093/mnras/stab3410},
   DOI={10.1093/mnras/stab3410},
   number={4},
   journal={Monthly Notices of the Royal Astronomical Society},
   publisher={Oxford University Press (OUP)},
   author={Fang, Xiao and Eifler, Tim and Schaan, Emmanuel and Huang, Hung-Jin and Krause, Elisabeth and Ferraro, Simone},
   year={2021},
   month=nov, pages={5721–5736} }

@article{Martinelli_2021,
   title={<i>Euclid</i>: Impact of non-linear and baryonic feedback prescriptions on cosmological parameter estimation from weak lensing cosmic shear},
   volume={649},
   ISSN={1432-0746},
   url={http://dx.doi.org/10.1051/0004-6361/202039835},
   DOI={10.1051/0004-6361/202039835},
   journal={Astronomy &amp; Astrophysics},
   publisher={EDP Sciences},
   author={Martinelli, M. and Tutusaus, I. and Archidiacono, M. and Camera, S. and Cardone, V. F. and Clesse, S. and Casas, S. and Casarini, L. and Mota, D. F. and Hoekstra, H. and Carbone, C. and Ilić, S. and Kitching, T. D. and Pettorino, V. and Pourtsidou, A. and Sakr, Z. and Sapone, D. and Auricchio, N. and Balestra, A. and Boucaud, A. and Branchini, E. and Brescia, M. and Capobianco, V. and Carretero, J. and Castellano, M. and Cavuoti, S. and Cimatti, A. and Cledassou, R. and Congedo, G. and Conselice, C. and Conversi, L. and Corcione, L. and Costille, A. and Douspis, M. and Dubath, F. and Dusini, S. and Fabbian, G. and Fosalba, P. and Frailis, M. and Franceschi, E. and Gillis, B. and Giocoli, C. and Grupp, F. and Guzzo, L. and Holmes, W. and Hormuth, F. and Jahnke, K. and Kermiche, S. and Kiessling, A. and Kilbinger, M. and Kunz, M. and Kurki-Suonio, H. and Ligori, S. and Lilje, P. B. and Lloro, I. and Maiorano, E. and Marggraf, O. and Markovic, K. and Massey, R. and Meneghetti, M. and Meylan, G. and Morin, B. and Moscardini, L. and Niemi, S. and Padilla, C. and Paltani, S. and Pasian, F. and Pedersen, K. and Pires, S. and Polenta, G. and Poncet, M. and Popa, L. and Raison, F. and Rhodes, J. and Roncarelli, M. and Rossetti, E. and Saglia, R. and Schneider, P. and Secroun, A. and Serrano, S. and Sirignano, C. and Sirri, G. and Starck, J.-L. and Sureau, F. and Taylor, A. N. and Tereno, I. and Toledo-Moreo, R. and Valentijn, E. A. and Valenziano, L. and Vassallo, T. and Wang, Y. and Welikala, N. and Zacchei, A. and Zoubian, J.},
   year={2021},
   month=May, pages={A100} }

@article{euclid_prep_XXXIV2023,
   title={Euclid preparation. XXXIV. The effect of linear redshift-space distortions in photometric galaxy clustering and its cross-correlation with cosmic shear},
   ISSN={1432-0746},
   url={http://dx.doi.org/10.1051/0004-6361/202347870},
   DOI={10.1051/0004-6361/202347870},
   journal={Astronomy &amp; Astrophysics},
   publisher={EDP Sciences},
   author={Tanidis, K. and Cardone, V.F. and Martinelli, M. and Tutusaus, I. and Camera, S. and Aghanim, N. and Amara, A. and Andreon, S. and Auricchio, N. and Baldi, M.},
   year={2023},
   month=dec }

@article{Lewis_2000,
   title={Efficient Computation of Cosmic Microwave Background Anisotropies in Closed Friedmann‐Robertson‐Walker Models},
   volume={538},
   ISSN={1538-4357},
   url={http://dx.doi.org/10.1086/309179},
   DOI={10.1086/309179},
   number={2},
   journal={The Astrophysical Journal},
   publisher={American Astronomical Society},
   author={Lewis, Antony and Challinor, Anthony and Lasenby, Anthony},
   year={2000},
   month=Aug, pages={473–476} }

@article{Pandey_2020,
   title={Perturbation theory for modeling galaxy bias: Validation with simulations of the Dark Energy Survey},
   volume={102},
   ISSN={2470-0029},
   url={http://dx.doi.org/10.1103/PhysRevD.102.123522},
   DOI={10.1103/physrevd.102.123522},
   number={12},
   journal={Physical Review D},
   publisher={American Physical Society (APS)},
   author={Pandey, S. and Krause, E. and Jain, B. and MacCrann, N. and Blazek, J. and Crocce, M. and DeRose, J. and Fang, X. and Ferrero, I. and Friedrich, O. and Aguena, M. and Allam, S. and Annis, J. and Avila, S. and Bernstein, G. M. and Brooks, D. and Burke, D. L. and Carnero Rosell, A. and Carrasco Kind, M. and Carretero, J. and Costanzi, M. and da Costa, L. N. and De Vicente, J. and Desai, S. and Elvin-Poole, J. and Everett, S. and Fosalba, P. and Frieman, J. and García-Bellido, J. and Gruen, D. and Gruendl, R. A. and Gschwend, J. and Gutierrez, G. and Honscheid, K. and Kuehn, K. and Kuropatkin, N. and Maia, M. A. G. and Marshall, J. L. and Menanteau, F. and Miquel, R. and Palmese, A. and Paz-Chinchón, F. and Plazas, A. A. and Roodman, A. and Sanchez, E. and Scarpine, V. and Schubnell, M. and Serrano, S. and Sevilla-Noarbe, I. and Smith, M. and Soares-Santos, M. and Suchyta, E. and Swanson, M. E. C. and Tarle, G. and Weller, J. and },
   year={2020},
   month=Dec }

@article{Mead_2021,
   title={<scp>hmcode-2020</scp>: improved modelling of non-linear cosmological power spectra with baryonic feedback},
   volume={502},
   ISSN={1365-2966},
   url={http://dx.doi.org/10.1093/mnras/stab082},
   DOI={10.1093/mnras/stab082},
   number={1},
   journal={Monthly Notices of the Royal Astronomical Society},
   publisher={Oxford University Press (OUP)},
   author={Mead, A J and Brieden, S and Tröster, T and Heymans, C},
   year={2021},
   month=jan, pages={1401–1422} }

@article{Smail_1994,
   title={Gravitational lensing of distant field galaxies by rich clusters - I. Faint galaxy redshift distributions},
   volume={270},
   ISSN={1365-2966},
   url={http://dx.doi.org/10.1093/mnras/270.2.245},
   DOI={10.1093/mnras/270.2.245},
   number={2},
   journal={Monthly Notices of the Royal Astronomical Society},
   publisher={Oxford University Press (OUP)},
   author={Smail, I. and Ellis, R. S. and Fitchett, M. J.},
   year={1994},
   month=sep, pages={245–270} }

@misc{xu2026,
      title={Constraining baryonic feedback and cosmology from DES Y3 and Planck PR4 6$\times$2pt data. I. $\Lambda$CDM models}, 
      author={Jiachuan Xu and Tim Eifler and Elisabeth Krause and Vivian Miranda and Jaime Salcido and Ian McCarthy},
      year={2026},
      eprint={2510.25596},
      archivePrefix={arXiv},
      primaryClass={astro-ph.CO},
      url={https://arxiv.org/abs/2510.25596}, 
}

@article{amy2025,
    author = {Wayland, Amy and Alonso, David and Zennaro, Matteo},
    title = {Calibrating baryonic effects in cosmic shear with external data in the LSST era},
    journal = {Monthly Notices of the Royal Astronomical Society},
    volume = {543},
    number = {2},
    pages = {1518-1534},
    year = {2025},
    month = {10},
    issn = {0035-8711},
    doi = {10.1093/mnras/staf1541},
    url = {https://doi.org/10.1093/mnras/staf1541},
    eprint = {https://academic.oup.com/mnras/article-pdf/543/2/1518/64262416/staf1541.pdf},
}

@article{Nelson_2015,
   title={The illustris simulation: Public data release},
   volume={13},
   ISSN={2213-1337},
   url={http://dx.doi.org/10.1016/j.ascom.2015.09.003},
   DOI={10.1016/j.ascom.2015.09.003},
   journal={Astronomy and Computing},
   publisher={Elsevier BV},
   author={Nelson, D. and Pillepich, A. and Genel, S. and Vogelsberger, M. and Springel, V. and Torrey, P. and Rodriguez-Gomez, V. and Sijacki, D. and Snyder, G.F. and Griffen, B. and Marinacci, F. and Blecha, L. and Sales, L. and Xu, D. and Hernquist, L.},
   year={2015},
   month=Nov, pages={12–37} }

@article{Lewis:2013hha,
      author         = "Lewis, Antony",
      title          = "{Efficient sampling of fast and slow cosmological
                        parameters}",
      journal   = "\prd",
      volume    = "87",
      year      = "2013",
      pages     = "103529",
      doi            = "10.1103/PhysRevD.87.103529",
      eprint         = "1304.4473",
      archivePrefix  = "arXiv",
      primaryClass   = "astro-ph.CO",
      SLACcitation   = "%%CITATION = ARXIV:1304.4473;%%",
}

@article{Seljak_2000,
   title={Analytic model for galaxy and dark matter clustering},
   volume={318},
   ISSN={1365-2966},
   url={http://dx.doi.org/10.1046/j.1365-8711.2000.03715.x},
   DOI={10.1046/j.1365-8711.2000.03715.x},
   number={1},
   journal={Monthly Notices of the Royal Astronomical Society},
   publisher={Oxford University Press (OUP)},
   author={Seljak, U.},
   year={2000},
   month=Oct, pages={203–213} }

@article{Fisher_1994,
   title={A spherical harmonic approach to redshift distortion and a measurement of Formula from the 1.2-Jy IRAS Redshift Survey},
   volume={266},
   ISSN={1365-2966},
   url={http://dx.doi.org/10.1093/mnras/266.1.219},
   DOI={10.1093/mnras/266.1.219},
   number={1},
   journal={Monthly Notices of the Royal Astronomical Society},
   publisher={Oxford University Press (OUP)},
   author={Fisher, K. B. and Scharf, C. A. and Lahav, O.},
   year={1994},
   month=jan, pages={219–226} }

@article{Padmanabhan_2007,
   title={The clustering of luminous red galaxies in the Sloan Digital Sky Survey imaging data},
   volume={378},
   ISSN={1365-2966},
   url={http://dx.doi.org/10.1111/j.1365-2966.2007.11593.x},
   DOI={10.1111/j.1365-2966.2007.11593.x},
   number={3},
   journal={Monthly Notices of the Royal Astronomical Society},
   publisher={Oxford University Press (OUP)},
   author={Padmanabhan, Nikhil and Schlegel, David J. and Seljak, Uroš and Makarov, Alexey and Bahcall, Neta A. and Blanton, Michael R. and Brinkmann, Jonathan and Eisenstein, Daniel J. and Finkbeiner, Douglas P. and Gunn, James E. and Hogg, David W. and Ivezić, Željko and Knapp, Gillian R. and Loveday, Jon and Lupton, Robert H. and Nichol, Robert C. and Schneider, Donald P. and Strauss, Michael A. and Tegmark, Max and York, Donald G.},
   year={2007},
   month=jun, pages={852–872} }

@article{Schaan_2020,
doi = {10.1088/1475-7516/2020/12/001},
url = {https://doi.org/10.1088/1475-7516/2020/12/001},
year = {2020},
month = {dec},
publisher = {},
volume = {2020},
number = {12},
pages = {001},
author = {Schaan, Emmanuel and Ferraro, Simone and Seljak, Uros},
title = {Photo-z outlier self-calibration in weak lensing surveys},
journal = {Journal of Cosmology and Astroparticle Physics}
}

@article{PhysRevD.100.103506,
  title = {Beyond linear galaxy alignments},
  author = {Blazek, Jonathan A. and MacCrann, Niall and Troxel, M. A. and Fang, Xiao},
  journal = {Phys. Rev. D},
  volume = {100},
  issue = {10},
  pages = {103506},
  numpages = {19},
  year = {2019},
  month = {Nov},
  publisher = {American Physical Society},
  doi = {10.1103/PhysRevD.100.103506},
  url = {https://link.aps.org/doi/10.1103/PhysRevD.100.103506}
}

@article{Guo_2013,
doi = {10.1088/0067-0049/207/2/24},
url = {https://doi.org/10.1088/0067-0049/207/2/24},
year = {2013},
month = {jul},
publisher = {The American Astronomical Society},
volume = {207},
number = {2},
pages = {24},
author = {Guo, Yicheng and Ferguson, Henry C. and Giavalisco, Mauro and Barro, Guillermo and Willner, S. P. and Ashby, Matthew L. N. and Dahlen, Tomas and Donley, Jennifer L. and Faber, Sandra M. and Fontana, Adriano and Galametz, Audrey and Grazian, Andrea and Huang, Kuang-Han and Kocevski, Dale D. and Koekemoer, Anton M. and Koo, David C. and McGrath, Elizabeth J. and Peth, Michael and Salvato, Mara and Wuyts, Stijn and Castellano, Marco and Cooray, Asantha R. and Dickinson, Mark E. and Dunlop, James S. and Fazio, G. G. and Gardner, Jonathan P. and Gawiser, Eric and Grogin, Norman A. and Hathi, Nimish P. and Hsu, Li-Ting and Lee, Kyoung-Soo and Lucas, Ray A. and Mobasher, Bahram and Nandra, Kirpal and Newman, Jeffery A. and van der Wel, Arjen},
title = {CANDELS MULTI-WAVELENGTH CATALOGS: SOURCE DETECTION AND PHOTOMETRY IN THE GOODS-SOUTH FIELD},
journal = {The Astrophysical Journal Supplement Series}
}

@article{Mandelbaum_2018,
   title={Weak Lensing for Precision Cosmology},
   volume={56},
   ISSN={1545-4282},
   url={http://dx.doi.org/10.1146/annurev-astro-081817-051928},
   DOI={10.1146/annurev-astro-081817-051928},
   number={1},
   journal={Annual Review of Astronomy and Astrophysics},
   publisher={Annual Reviews},
   author={Mandelbaum, Rachel},
   year={2018},
   month=Sept, pages={393–433} }

@article{Weinberg_2013,
   title={Observational probes of cosmic acceleration},
   volume={530},
   ISSN={0370-1573},
   url={http://dx.doi.org/10.1016/j.physrep.2013.05.001},
   DOI={10.1016/j.physrep.2013.05.001},
   number={2},
   journal={Physics Reports},
   publisher={Elsevier BV},
   author={Weinberg, David H. and Mortonson, Michael J. and Eisenstein, Daniel J. and Hirata, Christopher and Riess, Adam G. and Rozo, Eduardo},
   year={2013},
   month=Sept, pages={87–255} }

@article{2020euclidforecast,
   title={<i>Euclid</i> preparation: VII. Forecast validation for <i>Euclid</i> cosmological probes},
   volume={642},
   ISSN={1432-0746},
   url={http://dx.doi.org/10.1051/0004-6361/202038071},
   DOI={10.1051/0004-6361/202038071},
   journal={Astronomy \& Astrophysics},
   publisher={EDP Sciences},
   author={Euclid Collaboration and Blanchard, A. and Camera, S. and Carbone, C. and Cardone, V. F. and Casas, S. and Clesse, S. and Ilić, S. and Kilbinger, M. and Kitching, T. and Kunz, M. and Lacasa, F. and Linder, E. and Majerotto, E. and Markovič, K. and Martinelli, M. and Pettorino, V. and Pourtsidou, A. and Sakr, Z. and Sánchez, A. G. and Sapone, D. and Tutusaus, I. and Yahia-Cherif, S. and Yankelevich, V. and Andreon, S. and Aussel, H. and Balaguera-Antolínez, A. and Baldi, M. and Bardelli, S. and Bender, R. and Biviano, A. and Bonino, D. and Boucaud, A. and Bozzo, E. and Branchini, E. and Brau-Nogue, S. and Brescia, M. and Brinchmann, J. and Burigana, C. and Cabanac, R. and Capobianco, V. and Cappi, A. and Carretero, J. and Carvalho, C. S. and Casas, R. and Castander, F. J. and Castellano, M. and Cavuoti, S. and Cimatti, A. and Cledassou, R. and Colodro-Conde, C. and Congedo, G. and Conselice, C. J. and Conversi, L. and Copin, Y. and Corcione, L. and Coupon, J. and Courtois, H. M. and Cropper, M. and Da Silva, A. and de la Torre, S. and Di Ferdinando, D. and Dubath, F. and Ducret, F. and Duncan, C. A. J. and Dupac, X. and Dusini, S. and Fabbian, G. and Fabricius, M. and Farrens, S. and Fosalba, P. and Fotopoulou, S. and Fourmanoit, N. and Frailis, M. and Franceschi, E. and Franzetti, P. and Fumana, M. and Galeotta, S. and Gillard, W. and Gillis, B. and Giocoli, C. and Gómez-Alvarez, P. and Graciá-Carpio, J. and Grupp, F. and Guzzo, L. and Hoekstra, H. and Hormuth, F. and Israel, H. and Jahnke, K. and Keihanen, E. and Kermiche, S. and Kirkpatrick, C. C. and Kohley, R. and Kubik, B. and Kurki-Suonio, H. and Ligori, S. and Lilje, P. B. and Lloro, I. and Maino, D. and Maiorano, E. and Marggraf, O. and Martinet, N. and Marulli, F. and Massey, R. and Medinaceli, E. and Mei, S. and Mellier, Y. and Metcalf, B. and Metge, J. J. and Meylan, G. and Moresco, M. and Moscardini, L. and Munari, E. and Nichol, R. C. and Niemi, S. and Nucita, A. A. and Padilla, C. and Paltani, S. and Pasian, F. and Percival, W. J. and Pires, S. and Polenta, G. and Poncet, M. and Pozzetti, L. and Racca, G. D. and Raison, F. and Renzi, A. and Rhodes, J. and Romelli, E. and Roncarelli, M. and Rossetti, E. and Saglia, R. and Schneider, P. and Scottez, V. and Secroun, A. and Sirri, G. and Stanco, L. and Starck, J.-L. and Sureau, F. and Tallada-Crespí, P. and Tavagnacco, D. and Taylor, A. N. and Tenti, M. and Tereno, I. and Toledo-Moreo, R. and Torradeflot, F. and Valenziano, L. and Vassallo, T. and Verdoes Kleijn, G. A. and Viel, M. and Wang, Y. and Zacchei, A. and Zoubian, J. and Zucca, E.},
   year={2020},
   month=Oct, pages={A191} }

@article{Fang__2020,
   title={2D-FFTLog: efficient computation of real-space covariance matrices for galaxy clustering and weak lensing},
   volume={497},
   ISSN={1365-2966},
   url={http://dx.doi.org/10.1093/mnras/staa1726},
   DOI={10.1093/mnras/staa1726},
   number={3},
   journal={Monthly Notices of the Royal Astronomical Society},
   publisher={Oxford University Press (OUP)},
   author={Fang, Xiao and Eifler, Tim and Krause, Elisabeth},
   year={2020},
   month=June, pages={2699–2714} }

@Article{         harris2020array,
 title         = {Array programming with {NumPy}},
 author        = {Charles R. Harris and K. Jarrod Millman and St{\'{e}}fan J.
                 van der Walt and Ralf Gommers and Pauli Virtanen and David
                 Cournapeau and Eric Wieser and Julian Taylor and Sebastian
                 Berg and Nathaniel J. Smith and Robert Kern and Matti Picus
                 and Stephan Hoyer and Marten H. van Kerkwijk and Matthew
                 Brett and Allan Haldane and Jaime Fern{\'{a}}ndez del
                 R{\'{i}}o and Mark Wiebe and Pearu Peterson and Pierre
                 G{\'{e}}rard-Marchant and Kevin Sheppard and Tyler Reddy and
                 Warren Weckesser and Hameer Abbasi and Christoph Gohlke and
                 Travis E. Oliphant},
 year          = {2020},
 month         = sep,
 journal       = {Nature},
 volume        = {585},
 number        = {7825},
 pages         = {357--362},
 doi           = {10.1038/s41586-020-2649-2},
 publisher     = {Springer Science and Business Media {LLC}},
 url           = {https://doi.org/10.1038/s41586-020-2649-2}
}

@ARTICLE{2020SciPy-NMeth,
  author  = {Virtanen, Pauli and Gommers, Ralf and Oliphant, Travis E. and
            Haberland, Matt and Reddy, Tyler and Cournapeau, David and
            Burovski, Evgeni and Peterson, Pearu and Weckesser, Warren and
            Bright, Jonathan and {van der Walt}, St{\'e}fan J. and
            Brett, Matthew and Wilson, Joshua and Millman, K. Jarrod and
            Mayorov, Nikolay and Nelson, Andrew R. J. and Jones, Eric and
            Kern, Robert and Larson, Eric and Carey, C J and
            Polat, {\.I}lhan and Feng, Yu and Moore, Eric W. and
            {VanderPlas}, Jake and Laxalde, Denis and Perktold, Josef and
            Cimrman, Robert and Henriksen, Ian and Quintero, E. A. and
            Harris, Charles R. and Archibald, Anne M. and
            Ribeiro, Ant{\^o}nio H. and Pedregosa, Fabian and
            {van Mulbregt}, Paul and {SciPy 1.0 Contributors}},
  title   = {{{SciPy} 1.0: Fundamental Algorithms for Scientific
            Computing in Python}},
  journal = {Nature Methods},
  year    = {2020},
  volume  = {17},
  pages   = {261--272},
  adsurl  = {https://rdcu.be/b08Wh},
  doi     = {10.1038/s41592-019-0686-2},
}

@Article{Hunter:2007,
  Author    = {Hunter, J. D.},
  Title     = {Matplotlib: A 2D graphics environment},
  Journal   = {Computing in Science \& Engineering},
  Volume    = {9},
  Number    = {3},
  Pages     = {90--95},
  publisher = {IEEE COMPUTER SOC},
  doi       = {10.1109/MCSE.2007.55},
  year      = 2007
}

\appendix

\section{Real-space data vector comparison}
\label{sec:realDV}
\vspace{0.2cm}
We briefly discuss two details in the real-space code comparison here: the angular bin window function (or, bin-average) and the flat-sky/curved-sky transformation.
The former refers to the fact that when we compare with an actual measured data vector, we need to average the correlation function model over a given angular bin $[\theta_\mathrm{min}, \theta_\mathrm{max}]$ where the data is measured. The latter relates to the exact mathematical transformation from angular power spectra to correlation functions. 

\begin{itemize}
\item \textit{Treatment of averaging the model within angular bin:} In earlier cosmic shear analyses such as DES Y1 \citep{krause2017darkenergysurveyyear}, the model value for a given angular bin $[\theta_{\rm min}, \theta_{\rm max}]$ is approximated by the model value at some effective mean angular value instead of averaging over the angular bin,
\begin{equation}
    \bar{\theta} = \frac{2}{3}\frac{\theta_\mathrm{max}^3 - \theta_\mathrm{min}^3}{\theta_\mathrm{max}^2 - \theta_\mathrm{min}^2}.
    \label{eq:binave}
\end{equation}
This approximation is made to avoid the computationally expensive numerical average over angular bin and is verified to be sufficient given the DES-Y1 statistical uncertainties. However, as an analytic bin-averaging approach is employed in DES-Y3, as discussed in Section~\ref{sec:model-correlation function}, the computational expense is no longer a problem for MCMC. As such, we argue that the effective mean angular value should not be used. 
To demonstrate this, we set up both surveys using the source- and lens-galaxy samples in Figure~\ref{fig:nofz}, calculate the correlation functions using \textsc{CoCoA} with analytical bin-average while employing the effective mean angular value described by Equation~\ref{eq:binave} using \textsc{CCL}. The resulting $\Delta\chi^2$ of the data vector comparison is dramatic even under our fiducial scale cuts which is $\Delta\chi_{\mathrm{fid-cut}}^2(\xi^\pm, \gamma_t, w, 3\times2\mathrm{pt})_\mathrm{Roman} = (0.399,1.849,3.488,4.998)$ for Roman HLIS-Y5 and $\Delta\chi_{\mathrm{fid-cut}}^2(\xi^\pm, \gamma_t, w, 3\times2\mathrm{pt})_\mathrm{Rubin} = (0.591, 8.420, 20.653, 26.363)$ for Rubin LSST-Y10.

\item \textit{Flat-sky/Curved-sky transformation:} Transforming angular power spectra into correlation functions is the last step of data vector modeling in real space. One can assume the observed sky patch is a flat plane to make this transformation mathematically easier and computationally cheaper, because projecting a power spectrum into the curved sky involves sums over multipoles of angular power spectra weighted by the $\ell$-dependent Legendre functions, a procedure which was thought to be computationally expensive, while the flat-sky transformation only requires an integral weighted by the Bessel functions that can be computed easily. However, this is no longer a barrier for MCMC as long as we cache those Legendre functions properly.

We note that some previous work only implemented curved-sky transformation for galaxy-galaxy lensing and galaxy clustering, while leaving cosmic shear assuming flat-sky.  
We find that at least for Roman HLIS-Y5 and Rubin LSST-Y10, this approximation will be insufficient.
When comparing the cosmic shear data vector with and without curved-sky transformation under the fiducial settings for Roman and Rubin, we find 
$\Delta \chi^2(\xi^{\pm}, \xi_{\mathrm{fid}-\mathrm{cut}}^\pm)_\mathrm{Roman}=(3.522,0.374)$ and $\Delta \chi^2(\xi^{\pm}, \xi_{\mathrm{fid}-\mathrm{cut}}^\pm)_\mathrm{Rubin}=(6.005,0.970)$. This demonstrates that, even with our fiducial scale cuts, the $\Delta\chi^2$ introduced by flat-sky against curved-sky is non-negligible for both surveys. The effect will also be larger for more aggressive scale cuts.

\end{itemize}

\section{Lens samples in existing literature}
\label{sec:lens_sample}
\vspace{0.2cm}
\begin{figure}
  \centering
  \includegraphics[width=\linewidth]{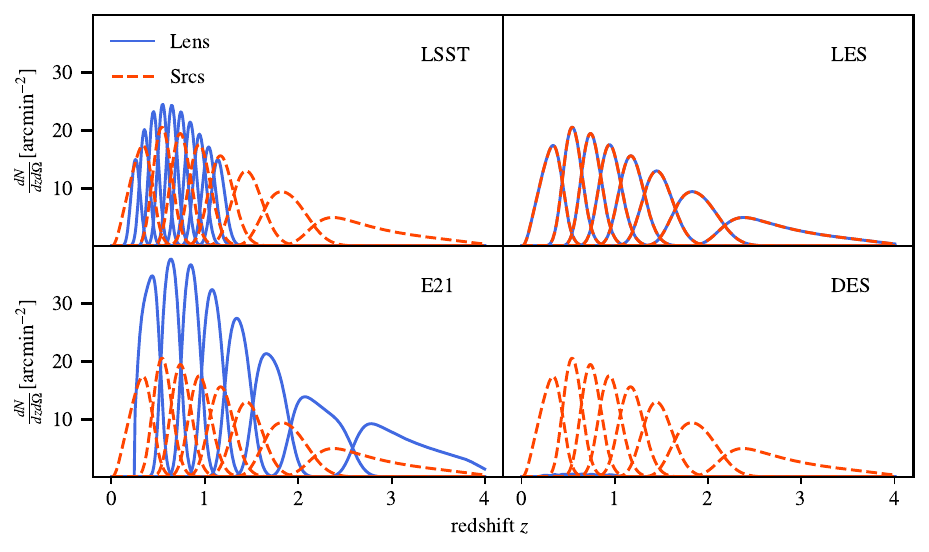}
  \caption{Lens and source redshift distributions of scenarios discussed in Appendix~\ref{sec:lens_sample}.
  }
  \label{fig:4nofz}
\end{figure}

As discussed in Section~\ref{sec:surveys}, there have been different assumptions for the Roman lens sample in the literature, while in this paper we choose to use yet a different sample that we deem more realistic. To assess the impact of this choice, we compare the constraining power for our fiducial Roman HLIS-Y5 survey when the different lens samples are used.
The four lens samples are:

\begin{itemize}
    \item \texttt{LSST}: This is the fiducial lens sample used in this paper. 
    The sample is defined using the Smail equation with the parameterization given by the LSST DESC SRD, then divided into 10 tomographic bins uniformly from redshift $0.2$ to $1.2$. The final effective number density for this lens sample is $n_\mathrm{eff} = 30\ \mathrm{arcmin}^{-2}$. 
    
    \item \texttt{LES}: A series of papers \citep{Fang_2021, Schaan_2020, boruah2024machinelearninglsst3x2pt,Cao2026} have directly adopted the source-galaxy sample as the lens-galaxy sample, which has been argued to be beneficial for statistical power by reducing the number of nuisance parameters marginalized. However, the plausibility of this method is still uncertain since it has not been used in real surveys. 
    
    \item \texttt{E21}: In \citet{Eifler_2021}, the authors employ the Galaxy Survey Exposure Time Calculator (ETC) with CANDELS catalog \citep{Guo_2013}, using a different selection function, to generate the lens and source sample. We replicate that lens sample  
    following the medium tier definition of Roman HLIS-Y5. We set exposure time $5\times 107$ second for each pixel and band pass for YJH filters of Roman telescope. Since lens sample does not need shape selection, we set the minimum signal-to-noise ratio (S/N) to 10, maximum ellipticity error $\sigma_e$ to 0.99, and minimum resolution factor $R$ to 0.01. We fit the selected samples with the Smail equation with a fixed $\beta=2$, generate the analytical lens sample distribution, then divide the $z>0.2$ part into 8 tomographic bins with equivalent number density and convolve them with a Gaussian kernel of $\sigma_z=0.05$. The resulting lens sample has an effective number density $n_\mathrm{eff}=67\ \mathrm{arcmin}^{-2}$. 
    
    \item \texttt{DES}: We assume a lens sample that has a similar selection function of DES-Y6 while keeping the overall shape of \texttt{LSST} lens redshift distribution. 
    This results in a total number density of $n_\mathrm{eff} = 0.633\ \mathrm{arcmin}^{-2}$. We regard this as a conservative scenario in the sense that we can already characterize a lens sample like this and no new methodologies are needed.
\end{itemize}
Figure~\ref{fig:4nofz} illustrates the redshift distribution and number density for each of the lens samples, compared to the fixed source sample. The effective number density for each sample is listed in Table~\ref{tab:FoM_lens_sample}. 

\begin{table}
  \centering
  \caption{This table shows FoMs of probes (columns 2-4) and the number densities of lens and source samples (columns 5-6) for scenarios discussed in Appendix~\ref{sec:lens_sample}. FoMs are calculated based on the area of $\Omega_\mathrm{m}-\sigma_8$ plane with 68\% credible region. The number density is in units of $\mathrm{arcmin}^{-2}$.}
  \label{tab:FoM_lens_sample}
  \begin{tabular}{lccccc}
    \toprule
    FoMs$\ |\ n_\mathrm{eff}$ & $3\times2\mathrm{pt}$ & cosmic shear & $2\times2\mathrm{pt}$ & lens & source\\
    \midrule
    \texttt{LSST}   & $2.9\times10^4$ & $8.0\times10^3$ & $1.4\times10^4$ & 30 & 41\\
    \texttt{LES}    & $4.5\times10^4$ & $8.0\times10^3$ & $2.9\times10^4$ & 41 & 41\\
    \texttt{E21}    & $3.1\times10^4$ & $8.0\times10^3$ & $1.6\times10^4$ & 67 & 41 \\
    \texttt{DES}    & $1.9\times10^4$ & $8.0\times10^3$ & $7.0\times10^3$ & 0.633 & 41\\
    \bottomrule
  \end{tabular}
  \\
\end{table}

Given these lens samples and the Roman source sample, we thus follow Section~\ref{sec:inference} to generate the covariance matrix for each scenario, then implement the Fisher analysis following \citet{Cao2026} with our fiducial scale cuts defined in Section~\ref{sec:scale_cut} to obtain an upper-bound estimation of constraining power. We follow Section~\ref{sec:approximation_impact} to calculate the FoM based on $\Omega_\mathrm{m}-\sigma_8$ plane for constraints obtained from the Fisher and the results are shown in Table~\ref{tab:FoM_lens_sample}.

The results show a general trend that the constraining power of $3\times2\mathrm{pt}$ and $2\times2\mathrm{pt}$ rises with the increase of the lens-sample number density. The exception is that the \texttt{LES} case has the largest FoM even though the number density is lower than that of \texttt{E21}.
This is because lens sample and source sample in \texttt{LES} share the same set of photo-z nuisance parameters thus fewer nuisance parameters need to be marginalized. 
We also note that even though number density of lens sample of \texttt{DES} is two orders of magnitude lower than those of other scenarios, its FoM of $3\times2\mathrm{pt}$ is only 39\% less than that of the \texttt{E21} case.
This demonstrates that constraining power has diminishing returns from the increase of lens-sample number density. However, increasing the lens-sample number density may reduce photo-z accuracy\citep{porredon2021}, so the balance between the gain of constraining power and the systematics control is important when increase the lens-sample number density.

Table~\ref{tab:FoM_lens_sample} also makes clear how the balance between cosmic shear and 2$\times$2pt shifts as a function of the lens-sample number density.
As the number density increases, constraining power of $2\times2\mathrm{pt}$ becomes dominant relative to that of cosmic shear. In the \texttt{LES} scenario, the FoM of $2\times2\mathrm{pt}$ is more than three times than that of cosmic shear. In this scenario, one would imagine that the detailed characterization of the lens sample will become more important.   

\section{Impact of scale cut choices}
\label{app:analysis_choice}
\vspace{0.2cm}
As discussed in Section~\ref{sec:scale_cut}, our analysis assumes a fiducial scale cut that is similar to what is adopted by DES: a minimum transverse comoving radius $(R_\mathrm{min}^{\gamma t}, R_\mathrm{min}^w) = (6\,\mathrm{Mpc}/h,8\,\mathrm{Mpc}/h)$ for galaxy-galaxy lensing and galaxy clustering, and a $\Delta \chi^2$ threshold for cosmic shear. This is different from most forecasting papers for both Roman and Rubin. In particular, many Roman and Rubin forecast papers adopt the scale cut choices used in the LSST DESC SRD, where $\ell_\mathrm{max}=3000$ for cosmic shear and $k_\mathrm{max} = 0.3h/\mathrm{Mpc}$ for galaxy clustering and galaxy-galaxy lensing. In this case, few cosmic-shear data points are discarded by this scale cuts which emphasizes the assumption that we will be able to model the information of cosmic shear on such small scales. The $k_\mathrm{max} = 0.3\, h/\mathrm{Mpc}$ is equivalent to a minimum transverse radius $R_\mathrm{min}=21\mathrm{Mpc}/h$ which is more conservative compared to the number used in DES, discarding about 1/2 data points from galaxy-galaxy lensing and galaxy clustering. Here we denote our fiducial scale cuts as \texttt{DES-like} while the SRD assumed scale cuts as \texttt{SRD-like}.

We also introduce two different lens samples, \texttt{LES} and \texttt{E21}, into this comparison of scale cuts, against our fiducial lens sample \texttt{LSST}. We thus study six different combinations of lens samples and scale cuts: \texttt{LSST}$\times$\texttt{DES-like}, \texttt{LES}$\times$\texttt{DES-like}, \texttt{E21}$\times$\texttt{DES-like}, \texttt{LSST}$\times$\texttt{SRD-like}, \texttt{LES}$\times$\texttt{SRD-like}, and \texttt{E21}$\times$\texttt{SRD-like}. The first scenario is exactly our baseline in this paper, while the last scenario is closer to previous Roman forecast papers.

\begin{figure}
  \centering
  \includegraphics[width=\textwidth]{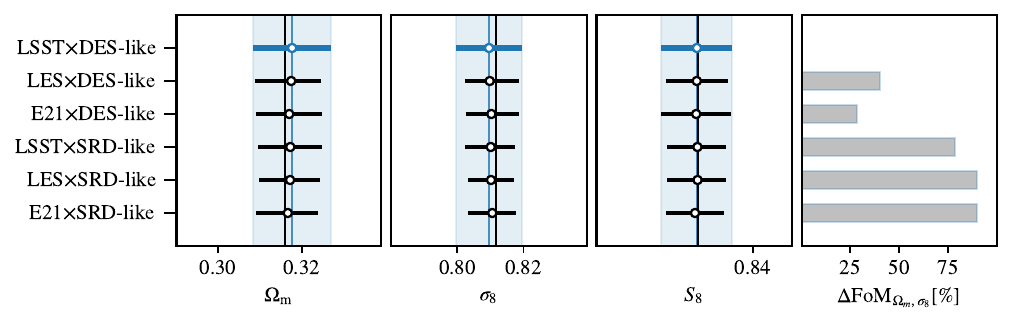}
  \caption{The left three panels compare the posterior mean of parameters $\sigma_8$, $\Omega_\mathrm{m}$, and $S_8$, under different scenarios which are explained in Appendix~\ref{app:analysis_choice}. The vertical black solid line represents the input fiducial, while the blue horizontal line is the posterior mean of our baseline, the difference between the two shows the level of projection effect. The rightmost plot shows the FoM defined in the plane of parameter pair $\sigma_8-\Omega_\mathrm{m}$ under different scenarios.}
  \label{fig:analysis-choice-FoM}
\end{figure}

We then run MCMC assuming the fiducial cosmology in this paper, and the marginal distributions of parameters $\Omega_\mathrm{m}$, $\sigma_8$, and $S_8$ are shown in Figure~\ref{fig:analysis-choice-FoM}. It is very clear that our baseline has the least constraining power and using the same galaxy sample for lens and source is beneficial to statistical power. The \texttt{SRD-like} scale cuts are the dominant factor boosting the constraining power as demonstrated at the rightmost panel of Figure~\ref{fig:analysis-choice-FoM}. The figure illustrates that neither of these scenarios leads to a noticeable inference bias though their constraining power is distinct, which can make different physical modeling effects have different priorities under different lens samples and scale cuts. For example, an effect that is ignorable in \texttt{LSST}$\times$ \texttt{DES-like} can be important in \texttt{LES}$\times$\texttt{SRD-like} due to the latter scenario having much stronger constraining power. This reveals the importance of having the most realistic galaxy sample and scale cuts before we determine how an effect is important in the analysis pipeline.

\end{document}